# Online adaptive group-wise sparse NPLS for ECoG neural signal decoding

Alexandre Moly[1*], Alexandre Aksenov[2], Alim Louis Benabid[1], Tetiana Aksenova[1*]

[1] Univ. Grenoble Alpes, CEA, LETI, Clinatec, F-38000 Grenoble
[2] Independent researcher
* Correspondence and requests for materials should be addressed to A.M. and T.A. (email: alexandre.moly@cea.fr and tetiana.aksenova@cea.fr).

## Abstract

*Objective*. Brain-computer interfaces (BCIs) create a new communication pathway between the brain and an effector without neuromuscular activation. BCI experiments highlighted high intra and inter-subjects variability in the BCI decoders. Although BCI model is generally relying on neurological markers generalizable on the majority of subjects, it requires to generate a wide range of neural features to include possible neurophysiological patterns. However, the processing of noisy and high dimensional features, such as brain signals, brings several challenges to overcome such as model calibration issues, model generalization and interpretation problems and hardware related obstacles.

*Approach*. An online adaptive group-wise sparse decoder named $L_p$-Penalized REW-NPLS algorithm (PREW-NPLS) is presented to reduce the feature space dimension employed for BCI decoding. The proposed decoder was designed to create BCI systems with low computational cost suited for portable applications and tested during offline pseudo-online study based on online closed-loop BCI control of the left and right 3D arm movements of a virtual avatar from the ECoG recordings of a tetraplegic patient.

*Main results*. PREW-NPLS algorithm highlight at least as good decoding performance as REW-NPLS algorithm. However, the decoding performance obtained with PREW-NPLS were achieved thanks to sparse models with up to 64% and 75% of the electrodes set to 0 for the left and right hand models respectively using $L_1$-PREW-NPLS. Additionally, $L_0$-PREW-NPLS and $L_1$-PREW-NPLS are computationally efficient and adapted for online closed-loop decoder adaptation at a high frequency rate (every tenth of seconds).

*Significance*. The designed solution proposed an online incremental adaptive algorithm suitable for online adaptive decoder calibration which estimate sparse decoding solutions. The PREW-NPLS models are suited for portable applications with low computational power using only small number of electrodes with degrading the decoding performance.

The studied data are extracted from a clinical trial catalogued in the ClinicalTrials.gov register, under the identifier: NCT02550522

## 1. Introduction

Brain-computer interfaces (BCIs) are systems that allow the control of external devices from the brain's neural signals without neuromuscular activation. Among the various applications, functional compensation and rehabilitation of individuals suffering from severe motor disabilities (with motor BCIs) has always been a focus for BCI research. With the exception of algorithms specifically oriented to this problematic, higher dimensional models require more training data set. Nevertheless, real-time BCI experiments are performed during rare and brief sessions due to the reduced ability of disabled patients to remain focused in prolonged calibration sequences [5]. Therefore, generally, calibration sessions are too short for complex high dimensional parameter identification and may lead to the classical "curse of dimensionality" problem related to uninformative or correlated features and small training dataset compared to feature space dimension [1–4]. Additionally, high dimensional feature spaces and related models with high number of parameters are more complicated to interpret than low dimensional one. Moreover, high dimensional feature spaces computation and high dimensional models evaluation require high computational power and time for neural signal processing, model calibration and application [1–4]. These hardware considerations are key characteristics in the case of real-time embedded/portable BCI application which have limited computational resources.

To prevent these issues, dimensional reduction algorithms decreasing the feature space dimension were employed to create the BCI model. Reduction of the feature space dimension may improve the decoding performance and drastically reduce the computing time and resources required. In the case of daily life BCI applications with high dimensional data flow processing, computing time and resources management is a crucial aspect to consider [6]. Dimensional reduction algorithms are dissociated into projection and feature selection algorithm families. Both dimensional reduction algorithm types were applied to BCI online experiments and offline studies.

**Projections algorithms** are often used in BCI studies [6–22]. They project the feature space into a subspace of lower dimension by linear or non-linear combination of the initial feature space components. This family clusters the principal and independent component analysis (PCA and ICA), spatio-spectral decomposition (SSD), common spatial pattern (CSP) or partial least squares (PLS) etc. [6–22]. However, such methods may not improve the computing time as they does not optimize feature extraction step. The irrelevant features are still computed.

**Feature selection family** regroups filter-based, wrapper-based and embedded techniques [23,24]. Filter-based methods ranks and selects independently the features which clusters most information without consideration of the trained decoder. This method is effective in computation time and have good generalization capacity. However, these methods tend to select highly correlated (redundant) features.

In the opposite, wrapper-based techniques incorporate supervised learning algorithms to evaluate the possible interactions between the features. Wrapper methods add iteratively new features to the subset of selected features space and evaluate the performance of the selected subset combined with the trained decoder [7]. These methods are efficient, nevertheless, they are costly in terms of computing time and may lead to overfitting.

Embedded techniques regroups the strategies were the feature selection steps is directly integrated into the decoding algorithm to combined the benefits of both previous methods: keeping the advantages of wrapper while decreasing computational complexity [23]. Features selection is performed directly within the model learning process. BCI Embedded techniques regroups decision tree, and regularization algorithms. Regularization strategies add penalty on the model parameter optimization to reduce the freedom of the model. Numerous regularization are named depending on the penalization term: $L_0$, $L_1$ (Lasso), $L_2$ (Ridge) or elastic net regularization algorithms etc. The $L_1$ regularization adds a penalty term equal to the sum of the absolute value of the coefficients whereas $L_2$ regularization integrates a penalty equal to the sum of the squared value of the coefficients and elastic net regularization is the combination of both $L_1$ and $L_2$ penalization [3]. $L_p$ regularization algorithms with $0 \leq p \leq 1$ discard irrelevant features promoting sparse solution [3,25]. Sparse solution is efficient to avoid overfitting and may lead to reduction in computing time.

Regularization algorithms were commonly applied in BCI field for feature selection or to improve neural signal decoding such as $L_0$ [26], $L_1$ [27–31], $L_2$ [13,27,32,33], elastic net [34,35] or other regularization strategies such as regularization using, polynomial regression [36], sparse regularization based on automatic relevance determination (ARD) parameters [37,38], Kullback-Leibler regularization in the Riemannian mean [39] etc. Generally, regularization algorithms is performed in single-wise manner, they evaluate independently the contribution of each model parameters before to apply constrain in order to regulate the amplitude of each parameter weights. Each features are regularized independently and are not evaluated as belonging to a group of features to be penalized. Therefore, in the case of tensor input features, each tensor components are set to zero independently to each other. Such element-wise component regularization of tensor features may lead to more complicated interpretation of the results and extraction of the relevant features. However, there are many applications with structurally grouped input features where it may be of interest to set simultaneously to zero or non-zero value features within a pre-determined group [25].

Group-wise regularization performs selection by grouping the relevant features and applying the penalization to the groups of features at once [25,30,31,40,41]. Grouping can clusters features over the electrodes, the frequency bands [42] etc. Group-wise sparse regularization promotes the model convergence to sparse solution (in a group-wise level), simplify the model interpretation and is suited to naturally structured features. Moreover, group-wise selection discards group of variables from the signal processing workflow (electrode or frequency) reducing the computation cost and the required computing time for real-time applications. Group-wise penalization was rarely applied to the BCI field [30,42–44]. Regularized PARAFAC and Tucker decomposition are two solutions designed for group-wise tensor penalization. Tensor is expressed as a linear combination of vectors which are independently regularized. Regularized tensor decomposition lead to a slice-wise tensor penalization creating more easily interpretable solution than element-wise regularization strategy. These approaches were used in some offline BCI experiments [30,41] and in other fields [40,45–47].

In BCI studies, most of the presented feature dimensional reduction algorithms were tested during offline experiment analysis [10–13,16,18–20,22,26–31,35,37–39,42–44,48–52]. Nevertheless, some of them were applied in online applications. Generally, feature selection was performed in offline preliminary studies before to apply the set of selected feature during online clinical or preclinical BCI experiments [8,14,33,34,53–56].

Online adaptive dimensional reduction strategies have advantages for online adaptive BCI. Majority of decoders trained in real-time are sensitive to overfitting due to lack of training data. Moreover, reduced feature space dimensions may reduce the required computing resources to apply the model in real time with faster data flow analysis.

The adaptive dimensional reduction algorithms commonly applied in the BCI [7,15,17,57–62] and other [63] fields were based on projections strategies such as adaptive CSP, PCA, ICA or xDAWN. However, all of them were only tested during offline studies. None of them were integrated into a BCI software made of an adaptive dimensional reduction procedure followed by an adaptive classifier/regression decoder.

Few adaptive feature selection algorithms have been designed in the motor BCI field. Filter methods have been tested on BCI simulation using Mutual Information [48] or during online BCI experiments based on Fisher score [64]. Wrapper strategy has been optimized using parallel computation for online BCI classifier [65] whereas embedded methods using semi-supervised feature selection [66] and weighting features algorithm [67] have been designed and used during online BCI applications. Adaptive genetic algorithm was proposed for adaptive channel selection in [68]. Nevertheless, all these algorithms were applied to simple online binary classification BCI experiments.

Regularized algorithms trained offline have been applied during online BCI experiments [32,69,70]. Adaptive regularized algorithms with fixed penalization hyperaparemeter were tested using offline dataset but none of these algorithms have been applied to real time BCI experiments [39,71,72]. Adaptive $L_1$ regularization strategy was applied to adaptive logistic regression [73], Kernel least squares [74] and recursive least squares algorithms [75] in other domains.

Only few dimensional reduction methods have been integrated into adaptive algorithms for online incremental calibration during real-time BCI experiments and are restricted to EEG-based experiments [64–68]. Computational complexity and difficulty to integrate dimensional reduction methods into real-time algorithms may explain the lack of solutions. Moreover, dimensional reduction methods often rely on hyperparameters which required to be tuned to optimize the decoding performances. This hyperparameter optimization problem may be another explanation of the lack of regularized adaptive decoder in the BCI fields.

In motor BCI field, $L_1$-Regularized N-way PLS algorithm developed by Eliseyev [30] and RPLS proposed by Foodeh [76] outperformed their non-penalized version due to noisy/non relevant electrodes suppression. However, these algorithms were not adapted to online adaptive decoding, required preliminary studies to fixed the hyperparameters and were only tested offline on non-human primates using ECoG [30,76] and Local field potential (LFP) on rats [76].

In the next section, the new Penalized REW-NPLS (PREW-NPLS) is proposed. PREW-NPLS is a new regularized recursive exponentially weighted N-way PLS designed for online adaptive decoding promoting group-wise (slice-wise) sparsity generalized to $L_0$, $L_{0,5}$ and $L_1$ norm regularization. PREW-NPLS rely on the REW-NPLS algorithms. The crucial REW-NPLS tensor decomposition procedure inspired from PARAFAC algorithm is modified to estimate a sparse $L_0$, $L_{0,5}$ or $L_1$ PARAFAC tensor decomposition. PREW-NPLS is an incremental adaptive regression algorithm which incrementally estimates a sparse $L_0$, $L_{0,5}$ and $L_1$ solution with a fixed penalization hyperparameter.

Firstly, in order to understand the proposed algorithms, the PARAFAC procedure employed in the non-penalized REW-NPLS algorithm is detailed [21]. Then, the new $L_p$-Penalized REW-NPLS algorithm (PREW-NPLS) for online sparse model identification is described. Finally, the results obtained in the offline pseudo-online studies carried out with PREW-NPLS algorithm for 3D left and right hand trajectory decoding from ECoG neural signals recorded during online closed-loop control of a virtual avatar by a tetraplegic are presented.

## 2. METHODS

### 2.1. PARAFAC procedure

REW-NPLS algorithm evaluates a set of projectors from the covariance matrix $\underline{\mathbf{XY}}_u$ using a rank one decomposition to evaluate the model parameters. Several tensor decomposition strategies were designed such as the PARAFAC, Tucker and multilinear SVD decomposition. The tensor decomposition employed in REW-NPLS algorithm is based on Parallel factor analysis (PARAFAC) tensor decomposition procedure. It is described in further detail in the next section.

#### 2.1.1. Definition

Parallel factor analysis (PARAFAC) or CANDECOMP/PARAFAC (CP) also known as polyadic decomposition (PD) can be considered as the generalization of principal component analysis (PCA) and singular value decomposition (SVD) to the tensor case [77,78]. This method represent a $M$-order tensor $\underline{\mathbf{V}} \in \mathbb{R}^{I_1 \times \dots \times I_M}$ as the linear combination of vectors outer products (rank-one tensors) such as :

$$\underline{\mathbf{V}} = \sum_{r=1}^{R} \rho_r \mathbf{w}_r^1 \circ \mathbf{w}_r^2 \circ \dots \circ \mathbf{w}_r^M + \underline{\mathbf{E}},$$
$$\text{with } r, m : \|\mathbf{w}_r^m\| = 1.$$

Here, $1 \leq m \leq M$ corresponds to the m$^{\text{th}}$ mode/dimension of the tensor variable, "∘" is the (vector) outer product of the decomposition factors (projectors) $\mathbf{w}_r^m \in \mathbb{R}^{I_m}$, $R \in \mathbb{N}$ is the fixed number of rank-one tensors used to decompose the original tensor variable, $\rho_r$ is weight associated to each rank-one tensor of the decomposition and $\underline{\mathbf{E}} \in \mathbb{R}^{I_1 \times \dots \times I_M}$ is the tensor of residuals [78].

#### 2.1.2. PARAFAC decomposition computation

Tensor decomposition is an appealing tool since the last twenty years in various fields (audio, image, video processing, biomedical applications,…) due to the rising of high dimensional data [77]. Nevertheless, no specific algorithm determining the rank of tensor has been defined [78]. Consequently, the number of rank-one tensor decomposition $R$ is set to a sub-optimal value [78]. Fixing $R$ leads to solve a low-rank approximation problem which is an ill-posed problem [79]. Numerous algorithms has been designed to locally solve the problem.

Most of the solutions can be regrouped into two families: direct methods regrouping Alternating least square (ALS), direct trilinear decomposition (DTLD), and iterative Non-least squares methods such as self-weighted alternating tri-linear decomposition (SWATLD) or alternating slice-wise diagonalization (ASD). Hessian and gradient based methods regroup Newton-based algorithms damped Gauss-Newton with compression (dGNc), positive matrix factorization for 3-way arrays (PMF3) and high-order singular value decomposition (HOSVD) [78,80]. No agreements on the best solution has been found on the literature but ALS seems to generally leads to good quality decomposition even though it is slower than numerous algorithms such as ASD [80–82].

Alternating least square (ALS) method is the most popular algorithm for PARAFAC decomposition [78,81] due to ease of implementation. Nevertheless this algorithm has many drawbacks. ALS method can be long to converge without guarantee of finding a global minimum [77,78,83,84]. Moreover, ALS is dependent on the initialization of the decomposition factors [78]. Several methods have been design to improve ALS performances depending on the decomposition quality, computing resources, computation time [77,81] such as Tikhonov regularization, maximum block improvement method [84], coupled-eigenvalue (CE) post-processing [79] etc.

The dGNc and PFM3 algorithm show better results than ALS in [80] but are more computationally expensive. CE post-processing improves the decomposition of truncated HOSVD and Sequential rank-one approximation (SeROA) presented in [83] highlights good results that should be compared to ALS. An interesting solution proposed in [80] could be to combine the different algorithms to exploit the benefits of each one. SWATLD algorithm could be used to initialize the decomposition factors of the rank-one tensor decomposition for PMF3, dGN or ALS algorithms before to apply CE post-processing [79,80]. However, there is no consensus on the advantages of the proposed alternative compared to ALS algorithms [78,82]

Next section is mainly focused on the most widespread ALS algorithm employed in the REW-NPLS for the PARAFAC tensor decomposition.

### 2.1.3. ALS based PARAFAC decomposition

Alternating least square (ALS) method optimizes one projector ($\mathbf{w}_r^m \in \mathbb{R}^{I_m}$) at a time and fixes the others[77–79]. In the next section, PARAFAC decomposition is considered in the specific case of three-order tensor decomposition to simplify the notation and to be closer to the BCI application presented in the next chapters. All the presented equations are generalizable to N-order tensor decomposition procedure.

Let $\underline{\mathbf{V}} \in \mathbb{R}^{I_1 \times I_2 \times I_3}$ be a third order tensor which undergoes PARAFAC decomposition. The aim is to find a tensor $\underline{\widehat{\mathbf{V}}} \in \mathbb{R}^{I_1 \times I_2 \times I_3}$ equal to the linear combination of $R \in \mathbb{N}$ outer product of three normalized projectors $\mathbf{w}_r^1 \in \mathbb{R}^{I_1}$, $\mathbf{w}_r^2 \in \mathbb{R}^{I_2}$, $\mathbf{w}_r^3 \in \mathbb{R}^{I_3}$ weighted with the coefficient $\rho_r \in \mathbb{R}$:

$$\min_{\underline{\widehat{X}}} \|\underline{\mathbf{V}} - \underline{\widehat{\mathbf{V}}}\|,$$

$$\underline{\widehat{\mathbf{V}}} = \sum_{r=1}^R \rho_r \mathbf{w}_r^1 \circ \mathbf{w}_r^2 \circ \mathbf{w}_r^3,$$

$$\|\mathbf{w}_r^1\| = \|\mathbf{w}_r^2\| = \|\mathbf{w}_r^3\| = 1.$$

The factor matrices refers to the concatenation of the decomposition factors $\mathbf{W}^1 \in \mathbb{R}^{I_1 \times R}$, $\mathbf{W}^2 \in \mathbb{R}^{I_2 \times R}$, $\mathbf{W}^3 \in \mathbb{R}^{I_3 \times R}$ with $\mathbf{W}^1 = [\mathbf{w}_1^1\ \mathbf{w}_2^1\ ...\ \mathbf{w}_R^1]$. From the factor matrices and the weighting vector $\boldsymbol{\rho} \in \mathbb{R}^R$, PARAFAC decomposition can be expressed with the unfolded tensor shape [78]:

$$\widehat{\mathbf{V}}_{(1)} = \mathbf{W}^1 \boldsymbol{\rho} (\mathbf{W}^3 \odot \mathbf{W}^2)^T,$$
$$\widehat{\mathbf{V}}_{(2)} = \mathbf{W}^2 \boldsymbol{\rho} (\mathbf{W}^3 \odot \mathbf{W}^1)^T,$$
$$\widehat{\mathbf{V}}_{(3)} = \mathbf{W}^3 \boldsymbol{\rho} (\mathbf{W}^2 \odot \mathbf{W}^1)^T.$$

**The ALS is an iterative procedure which reduces the optimization problem to smaller sub-problem [82]. Each step of the ALS solves a linear regression problem with one vector feature. At each step ALS fixes two of the three matrices $\mathbf{W}^1$, $\mathbf{W}^2$ and $\mathbf{W}^3$ and reduce the problem to a linear least-squares optimization. E.g. it fixes $\mathbf{W}^2$ and $\mathbf{W}^3$ to solve $\mathbf{W}^1$ then solve $\mathbf{W}^2$ fixing $\mathbf{W}^1$ and $\mathbf{W}^3$ and, finally, the same operation is realized for $\mathbf{W}^3$. Firstly, $\mathbf{W}^2$ and $\mathbf{W}^3$ are fixed which leads to**

$$\min_{\widehat{\mathbf{W}}^1} \|\mathbf{V}_{(1)} - \widehat{\mathbf{W}}_{\boldsymbol{\rho}}^1 (\mathbf{W}^3 \odot \mathbf{W}^2)^T\|,$$

where $\widehat{\mathbf{W}}^1 \in \mathbb{R}^{I_1 \times R}$ is the estimated factor matrix following the first decomposition dimension with [78]:

$$\widehat{\mathbf{W}}_{\boldsymbol{\rho}}^1 = \widehat{\mathbf{W}}^1 \boldsymbol{\rho}. \tag{2.1.1}$$

Minimum is achieved for

$$\widehat{\mathbf{W}}_{\boldsymbol{\rho}}^1 = \mathbf{V}_{(1)} [(\mathbf{W}^3 \odot \mathbf{W}^2)^T]^\dagger,$$

which simplifies due to the Khatri-Rao pseudoinverse properties [78] to

$$\widehat{\mathbf{W}}_{\boldsymbol{\rho}}^1 = \mathbf{V}_{(1)} (\mathbf{W}^3 \odot \mathbf{W}^2)(\mathbf{W}^{3T} \mathbf{W}^3 * \mathbf{W}^{2T} \mathbf{W}^2)^\dagger.$$

$\mathbf{W}^2$ and $\mathbf{W}^3$ are estimated following the same steps by fixing $\mathbf{W}^1 = \widehat{\mathbf{W}}^1$ using column-wise normalization with (2.1.1) resulting in:

$$\widehat{\mathbf{W}}^1 = \mathbf{V}_{(1)} (\mathbf{W}^3 \odot \mathbf{W}^2) \left(\mathbf{W}^{3T} \mathbf{W}^3 * \mathbf{W}^{2T} \mathbf{W}^2\right)^\dagger,$$
$$\widehat{\mathbf{W}}^2 = \mathbf{V}_{(2)} (\mathbf{W}^3 \odot \mathbf{W}^1) \left(\mathbf{W}^{3T} \mathbf{W}^3 * \mathbf{W}^{1T} \mathbf{W}^1\right)^\dagger,$$
$$\widehat{\mathbf{W}}^3 = \mathbf{V}_{(3)} (\mathbf{W}^2 \odot \mathbf{W}^1) \left(\mathbf{W}^{2T} \mathbf{W}^2 * \mathbf{W}^{1T} \mathbf{W}^1\right)^\dagger.$$

This procedure is repeated until a specified condition is reached (fixed number of iteration, convergence criterion, etc.). The factor matrices are initialized with random values, by values estimated in previous iteration of the ALS algorithm or applying another algorithm such as DTLD [21,78,81,82].

### 2.1.4. PARAFAC decomposition in the REW-NPLS algorithm.

REW-NPLS algorithm uses PARAFAC-based decomposition to extract the set of projectors of $\underline{\mathbf{XY}}_u$. In the next section, the PARAFAC decomposition problem of the REW-NPLS algorithm is considered in the specific case of three order tensor decomposition $\underline{\mathbf{XY}}_u \in \mathbb{R}^{I_1 \times I_2 \times I_3}$, $\|\underline{\mathbf{XY}}_u\| = 1$ to simplify the notation and to be closer to the BCI application considered in the PhD thesis. Nevertheless, all the presented results can be generalized to the n order tensor decomposition.

At each iteration ($f$ is current iteration number) of REW-NPLS algorithm, one iteration of PARAFAC algorithm is used (rank one approximation, $R = 1$) to decompose tensor $\underline{\mathbf{XY}}_u$ and to estimate projectors $\mathbf{w}_f^1$, $\mathbf{w}_f^2$, $\mathbf{w}_f^3$:

$$\min_{\underline{\widehat{\mathbf{XY}}}_u} \|\underline{\mathbf{XY}}_u - \underline{\widehat{\mathbf{XY}}}_u\|,$$

$$\underline{\widehat{\mathbf{XY}}}_u = \rho_f \mathbf{w}_f^1 \circ \mathbf{w}_f^2 \circ \mathbf{w}_f^3,$$

$$\|\mathbf{w}_f^1\| = \|\mathbf{w}_f^2\| = \|\mathbf{w}_f^3\| = 1.$$

where $\|\cdot\|$, as a reminder, always referred to $L_2$ norm (Frobenius, Euclidian norm depending on the variable dimensions). Equally

$$\min_{\underline{\widehat{\mathbf{XY}}}_u} \|\underline{\mathbf{XY}}_u - \underline{\widehat{\mathbf{XY}}}_u\|^2 \tag{2.1.2}$$

$$\underline{\widehat{\mathbf{XY}}}_u = \rho_f \mathbf{w}_f^1 \circ \mathbf{w}_f^2 \circ \mathbf{w}_f^3,$$

$$\|\mathbf{w}_f^1\| = \|\mathbf{w}_f^2\| = \|\mathbf{w}_f^3\| = 1.$$

As in current paragraph only one iteration of REW-NPLS algorithm is considered, iteration index $f$ is discarded in current paragraph for the purpose of simplification.

This problem is no longer an ill-posed problem [83]. ALS algorithm guarantees to converge [85].

In the REW-NPLS algorithm, PARAFAC decomposition is solved using ALS algorithm [21]. It optimizes sequentially

$$\min_{\mathbf{w}^1} \left\| \underline{\mathbf{XY}}_{u_{(1)}} - \mathbf{w}^1 (\mathbf{w}^3 \otimes \mathbf{w}^2)^T \right\|^2, \tag{2.1.3}$$

$$\min_{\mathbf{w}^2} \left\| \underline{\mathbf{XY}}_{u_{(2)}} - \mathbf{w}^2 (\mathbf{w}^3 \otimes \mathbf{w}^1)^T \right\|^2, \tag{2.1.4}$$

$$\min_{\mathbf{w}^3} \left\| \underline{\mathbf{XY}}_{u_{(3)}} - \mathbf{w}^3 (\mathbf{w}^2 \otimes \mathbf{w}^1)^T \right\|^2 \tag{2.1.5}$$

until convergence [86]. In the case of three-order tensor, Least Square (LS) solutions for each step are expressed:

$$\mathbf{w}_\rho^1 = \underline{\mathbf{XY}}_{u_{(1)}} (\mathbf{w}^3 \odot \mathbf{w}^2)(\mathbf{w}^{3^T}\mathbf{w}^3 * \mathbf{w}^{2^T}\mathbf{w}^2)^\dagger.$$

As $\mathbf{w}^i \in \mathbb{R}^{I_i}$, the solution can be simplified using:

$$\left(\mathbf{w}_f^{3^T}\mathbf{w}_f^3 * \mathbf{w}_f^{2^T}\mathbf{w}_f^2\right) = \|\widehat{\mathbf{w}}_f^2\|^2 * \|\widehat{\mathbf{w}}_f^3\|^2 = \|\mathbf{w}_f^3 \otimes \mathbf{w}_f^2\|^2 \in \mathbb{R},$$

$$\text{and } \left(\mathbf{w}_f^3 \odot \mathbf{w}_f^2\right) = \left(\mathbf{w}_f^3 \otimes \mathbf{w}_f^2\right).$$

To obtain the solutions:

$$\mathbf{w}_\rho^1 = \frac{\underline{\mathbf{XY}}_{u_{(1)}}(\mathbf{w}^3 \otimes \mathbf{w}^2)}{\|\mathbf{w}^3 \otimes \mathbf{w}^2\|^2}. \tag{2.1.6}$$

Normalization allows the estimation of parameter $\rho_f$ and $\mathbf{w}^1$ with $\|\mathbf{w}^1\| = 1$. The same procedure is repeated to evaluate both $\mathbf{w}^2$ and $\mathbf{w}^3$ :

$$\mathbf{w}_\rho^2 = \frac{\underline{\mathbf{XY}}_{u_{(2)}}(\mathbf{w}^3 \otimes \mathbf{w}^1)}{\|\mathbf{w}^3 \otimes \mathbf{w}^1\|^2}, \tag{2.1.7}$$

$$\mathbf{w}_\rho^3 = \frac{\underline{\mathbf{XY}}_{u_{(3)}}(\mathbf{w}^2 \otimes \mathbf{w}^1)}{\|\mathbf{w}^2 \otimes \mathbf{w}^1\|^2}. \tag{2.1.8}$$

Each one is normalized to evaluate $\rho$ and $\mathbf{w}^1, \mathbf{w}^2, \mathbf{w}^3$ with $\|\mathbf{w}^1\| = \|\mathbf{w}^2\| = \|\mathbf{w}^3\| = 1$. These three solutions are successively computed until a convergence or maximum iteration number criterion is reached.

### 2.2. Lp-Penalized REW-NPLS

PREW-NPLS algorithm exploited a penalized version of the PARAFAC algorithm to create group-wise sparse solution. This algorithm is an online adaptive algorithm which introduced $L_p$ penalization with p being the classic lasso regularization ($L_1$) or less conventional penalization type such as $L_0$ and $L_{0.5}$. This section describes the penalized PARAFAC procedure and its integration into the REW-NPLS algorithm to build the new online adaptive sparse PREW-NPLS algorithm.

#### 2.2.1. Penalized PARAFAC procedure

In the PARAFAC-based algorithm used in REW-NPLS, ALS strategy fixes all projectors except one at each step of the algorithm. Consequently, each step of the ALS solved a linear regression with one vector feature. In this section, $L_0$, $L_{0,5}$ and $L_1$ regularized linear regression are simplified to be applied in online PARAFAC subroutine of REW-NPLS. The following equation will be presented in the case of three-order tensor and rank one ($R = 1$) PARAFAC decomposition to simplify the notations but can be generalized to N-order tensor.

Given a three order-tensor $\underline{\mathbf{V}} \in \mathbb{R}^{I_1 \times I_2 \times I_3}$ to decompose using regularized PARAFAC with ALS strategy and $\mathbf{w}^i \in \mathbb{R}^{*I_i}$ with $i = 1,2,3$ the decomposition factors estimated by the PARAFAC. Let us consider the unfolded tensor $\underline{\mathbf{V}}_{(i)}$ with $\underline{\mathbf{V}}_{(i)} = (\mathbf{v}_1^1| \ldots |\mathbf{v}_1^{I_1}) \in \mathbf{R}^{I_1 \times I_2 I_3}$ where $\mathbf{v}_i^j$ are the rows of matrix $\underline{\mathbf{V}}_{(i)}$. Taking into account that $(\mathbf{w}^2 \otimes \mathbf{w}^1)^T \in \mathbf{R}^{I_1 I_2}$, $(\mathbf{w}^3 \otimes \mathbf{w}^1)^T \in \mathbf{R}^{I_1 I_3}$ and $(\mathbf{w}^3 \otimes \mathbf{w}^2)^T \in \mathbf{R}^{I_2 I_3}$ are vectors, optimization tasks (2.1.3)-(2.1.5) are separated into element-wise optimization:

$$\min_{w_j^1}\|\mathbf{v}_1^j - w_j^1(\mathbf{w}^3 \otimes \mathbf{w}^2)^T\|^2 \quad j = 1, \ldots, I_1, \quad (2.2.1)$$

$$\min_{w_j^2}\|\mathbf{v}_2^j - w_j^2(\mathbf{w}^3 \otimes \mathbf{w}^1)^T\|^2 \quad j = 1, \ldots, I_2, \quad (2.2.2)$$

$$\min_{w_j^3}\|\mathbf{v}_3^j - w_j^3(\mathbf{w}^2 \otimes \mathbf{w}^1)^T\|^2 \quad j = 1, \ldots, I_3. \quad (2.2.3)$$

where $w_j^i$ are the projector's elements of vectors $\mathbf{w}^1 = (w_1^1, \ldots, w_{I_1}^1)^T \in \mathbb{R}^{*I_1}$, $\mathbf{w}^2 = (w_1^2, \ldots, w_{I_2}^2)^T \in \mathbb{R}^{*I_2}$, and $\mathbf{w}^3 = (w_1^3, \ldots, w_{I_3}^3)^T \in \mathbb{R}^{*I_3}$ estimated by the PARAFAC. (2.1.6)-(2.1.8) least square (LS) solution may be written as

$$(w_j^1)_{LS} = \frac{\mathbf{v}_1^j(\mathbf{w}^3 \otimes \mathbf{w}^2)}{\|\mathbf{w}^3 \otimes \mathbf{w}^2\|^2}, \quad j = 1, \ldots, I_1, \quad (2.2.4)$$

$$(w_j^2)_{LS} = \frac{\mathbf{v}_2^j(\mathbf{w}^3 \otimes \mathbf{w}^1)}{\|\mathbf{w}^3 \otimes \mathbf{w}^1\|^2}, \quad j = 1, \ldots, I_2 \quad (2.2.5)$$

$$(w_j^3)_{LS} = \frac{\mathbf{v}_3^j(\mathbf{w}^2 \otimes \mathbf{w}^1)}{\|\mathbf{w}^2 \otimes \mathbf{w}^1\|^2}, \quad j = 1, \ldots, I_3. \quad (2.2.6)$$

Sparse promoting penalization with protection using $L_p$ ($p = 0, \frac{1}{2}, 1$) norm/pseudo norms is proposed to be integrated to the cost function of REW-NPLS procedure to provide a group-wise sparse solutions, namely, solutions sparse by slices. Optimization task (2.1.2) is replaced by the optimization of the cost function penalized with $L_p$ ($p = 0, \frac{1}{2}, 1$) norm/pseudo norms.

$$\min\|\underline{\mathbf{V}} - \hat{\underline{\mathbf{V}}}\|^2 + P(\mathbf{w}^1, \mathbf{w}^2, \mathbf{w}^3), \quad (2.2.7)$$

$$P(\mathbf{w}^1, \mathbf{w}^2, \mathbf{w}^3) = \lambda_1 \|\mathbf{w}^1\|_{q,\mathcal{L}_1} + \lambda_2 \|\mathbf{w}^2\|_{q,\mathcal{L}_2} + \lambda_3 \|\mathbf{w}^3\|_{q,\mathcal{L}_3},$$

$$\|\mathbf{w}^1\| = \|\mathbf{w}^2\| = \|\mathbf{w}^3\| = 1.$$

Where $\|\mathbf{w}^i\|_{p,\mathcal{L}_i}$ for $p = 0, \frac{1}{2}, 1$ and $i = 1, 2, 3$ is denoted as :

$$\|\mathbf{w}^i\|_{0,\mathcal{L}_i} = \sum_{k \in \mathcal{L}_i}\left(1 - \delta_{0,w_k^i}\right),$$

$$\|\mathbf{w}^i\|_{1,\mathcal{L}_i} = \sum_{k \in \mathcal{L}_i} |w_k^i|,$$

$$\|\mathbf{w}^i\|_{\frac{1}{2},\mathcal{L}_i} = \sum_{k \in \mathcal{L}_i} \sqrt{|w_k^i|}.$$

Here, the regularization functions may only regularize a part of the indices (projector's elements) defined by a set $\mathcal{L}_i \subset \{1, 2, \ldots, I_i\}$ with $i = 1, 2, 3$ and protecting other elements of vector. $\mathcal{L}_i$ may vary depending on Rew NPLS iteration. $0 < \lambda_i \leq 1$ are regularization coefficients. The Kronecker delta $\delta_{0,w_k^i} = 1$ if $w_k^i = 0$, $\delta_{0,w_k^i} = 0$ otherwise.

The same ALS strategy (2.1.3)-(2.1.5) than the procedure used in conventional REW- NPLS is proposed to be applied for optimization (2.2.7). ALS fixed all projectors except one at each step of the algorithm, leading to the three successive optimization tasks:

$$\min_{\mathbf{w}^1} \left( \|\underline{\mathbf{V}}_{(1)} - \mathbf{w}^1(\mathbf{w}^3 \otimes \mathbf{w}^2)^T\|^2 + \lambda_1 \|\mathbf{w}^1\|_{q,\mathcal{L}_1} \right),$$

$$\min_{\mathbf{w}^2} \left( \|\underline{\mathbf{V}}_{(2)} - \mathbf{w}^2(\mathbf{w}^3 \otimes \mathbf{w}^1)^T\|^2 + \lambda_2 \|\mathbf{w}^2\|_{q,\mathcal{L}_2} \right),$$

$$\min_{\mathbf{w}^3} \left( \|\underline{\mathbf{V}}_{(3)} - \mathbf{w}^3(\mathbf{w}^2 \otimes \mathbf{w}^1)^T\|^2 + \lambda_3 \|\mathbf{w}^3\|_{q,\mathcal{L}_3} \right).$$

The solutions of non-regularized problem (2.2.4)-(2.2.6) are used as initial approximation and are referred as the Least Square (LS) solution noted $\mathbf{w}_{LS}^i$.

Previously, similar penalized ALS was considered in [30]. However the study was limited to $L_1$-norm and did not consider additional protection variables $\mathcal{L}_i$. Moreover, the problem was solved using non-adaptive NPLS regression for offline classification preclinical experiments and highlighted non-viable solution for real-time processing if more than 10 electrodes were considered [30]. In the current article, more general case of $L_p$ ($p = 0, \frac{1}{2}, 1$)-norm/pseudo-norm penalization with possible variable protection procedure is proposed and an efficient integration to REW-NPLS algorithm is carried out. Unlike the non- regularized ALS algorithm (2.1.3)-(2.1.5), norms of projectors are not arbitrary parameters any more due to penalization terms. Therefore, the normalization of current estimate is added into to ALS optimization cycle.

$$\min_{\widetilde{\mathbf{w}}^1} \left( \|\underline{\mathbf{V}}_{(1)} - \widetilde{\mathbf{w}}^1(\mathbf{w}^3 \otimes \mathbf{w}^2)^T\|^2 + \lambda_1 \|\widetilde{\mathbf{w}}^1\|_{q,\mathcal{L}_1} \right) \text{ with } \mathbf{w}^1 = \widetilde{\mathbf{w}}^1 / \|\widetilde{\mathbf{w}}^1\| \qquad (2.2.8)$$

$$\min_{\widetilde{\mathbf{w}}^2} \left( \|\underline{\mathbf{V}}_{(2)} - \widetilde{\mathbf{w}}^2(\mathbf{w}^3 \otimes \mathbf{w}^1)^T\|^2 + \lambda_2 \|\widetilde{\mathbf{w}}^2\|_{q,\mathcal{L}_2} \right) \text{ with } \mathbf{w}^2 = \widetilde{\mathbf{w}}^2 / \|\widetilde{\mathbf{w}}^2\| \qquad (2.2.9)$$

$$\min_{\widetilde{\mathbf{w}}^3} \left( \|\underline{\mathbf{V}}_{(3)} - \widetilde{\mathbf{w}}^3(\mathbf{w}^2 \otimes \mathbf{w}^1)^T\|^2 + \lambda_3 \|\widetilde{\mathbf{w}}^3\|_{q,\mathcal{L}_3} \right) \text{ with } \mathbf{w}^3 = \widetilde{\mathbf{w}}^3 / \|\widetilde{\mathbf{w}}^3\| \qquad (2.2.10)$$

Next, for faster computing, it can be noted that all considered regularization functions are decomposed as a sum of element-wise functions. Consequently, similarly to (2.2.2)-(2.2.4) optimization tasks (2.2.8)-(2.2.10) are split into element-wise optimization:

$$\min_{w_j^1} \left( \|\mathbf{v}_1^j - w_j^1(\mathbf{w}^3 \otimes \mathbf{w}^2)^T\|^2 + \lambda_1 g_p(w_j^1) \right), j = 1, \ldots, I_1 \qquad (2.2.11)$$

$$\min_{w_j^2} \left( \|\mathbf{v}_2^j - w_j^2(\mathbf{w}^3 \otimes \mathbf{w}^1)^T\|^2 + \lambda_2 g_p(w_j^2) \right), j = 1, \ldots, I_2 \qquad (2.2.12)$$

$$\min_{w_j^3} \left( \|\mathbf{v}_3^j - w_j^3(\mathbf{w}^2 \otimes \mathbf{w}^1)^T\|^2 + \lambda_3 g_p(w_j^3) \right), j = 1, \ldots, I_3 \qquad (2.2.13)$$

$$g_p(w_j^i) = \begin{cases} 1 - \delta_{0,w_j^i}, & \text{if } p = 0 \text{ and } w_j^i \in \mathcal{L}_i \\ |w_j^i|, & \text{if } p = 1 \text{ and } w_j^i \in \mathcal{L}_i \\ \sqrt{|w_j^i|}, & \text{if } p = 1/2 \text{ and } w_j^i \in \mathcal{L}_i \\ 0 & \text{otherwise} \end{cases} \quad (2.2.14)$$

In the next subsections, the particular cases of L$_0$, L$_1$, L$_{1/2}$ penalizations are presented. Details of the demonstration are available in the Supplementary Materials.

**In the case of L$_0$ penalization** which penalized the parameter weights depending on the number of non-zero coefficients, and considering one of the optimization step, e.g. (2.2.11) of ALS optimization The solution turns out to be an element-wise hard thresholding of the least square solution $(w_j^1)_{LS}$ $j = 1, \ldots, I_1$ leading to:

$$(w_j^1)_{L_0} = \begin{cases} 0 & \text{if } j \in \mathcal{L}_1 \text{ and } (w_j^1)_{LS} \leq Threshold_{L_0} \\ (w_j^1)_{LS} & \text{otherwise} \end{cases},$$

$$Threshold_{L_0} = \frac{\sqrt{\lambda_1}}{\|\mathbf{w}^3 \otimes \mathbf{w}^2\|}.$$

**In the case of L$_{1/2}$ penalization** and considering one of the optimization step, e.g. (2.2.11) of ALS optimization. Based on (2.2.11) and (2.2.14), the function to minimize $FopL_{0.5}$ takes the form:

$$FopL_{0.5}(w_j^1) = \|\mathbf{v}_1^j - w_j^1(\mathbf{w}^3 \otimes \mathbf{w}^2)^T\|^2 + \lambda_1 \sqrt{|w_j^1|}, \quad (2.2.15)$$

or equivalently:

$$FopL_{0.5}(w_j^1) = \|\mathbf{w}^3 \otimes \mathbf{w}^2\|^2 \left((w_j^1)_{LS} - w_j^1\right)^2 + \lambda_1 \sqrt{|w_j^1|}. \quad (2.2.16)$$

The solution to this minimization problem is:

$$(w_j^1)_{L_{0.5}} = \begin{cases} 0, & \text{if } j \in \mathcal{L}_1 \text{ and } (w_j^1)_{LS} \leq ThresholdL_{0.5} \\ \operatorname{argmin}\left(FopL_{0.5}(0), FopL_{0.5}\left(\mathcal{B} \cdot (w_j^1)_{LS}\right)\right), & \text{if } i \in \mathcal{L}_1 \text{ and } (w_j^1)_{LS} > ThresholdL_{0.5}, \\ (w_j^1)_{LS} & \text{otherwise} \end{cases}$$

where

$$ThresholdL_{0.5} = \frac{3}{4}\left(\frac{\lambda_1}{\|\mathbf{w}^3 \otimes \mathbf{w}^2\|^2}\right)^{\frac{2}{3}},$$

and $\mathcal{B}$ is the solution of the cubic polynomial function (see figure S8 of the function in Supplementary Data):

$$x(1-x)^2 = C \quad (2.2.17)$$

$$\text{with } x = \frac{w_j^1}{(w_j^1)_{LS}} \text{ and } C = \frac{\lambda_1^2}{16\|\mathbf{w}^3 \otimes \mathbf{w}^2\|^4 \left((w_j^1)_{LS}\right)^3}.$$

To summarize, in the case $C > \frac{4}{27}$, $(w_j^1)_{L_{1/2}} = 0$ whereas in the case $C \in \left[0, \frac{4}{27}\right]$, By the properties of the Cubic polynomial function (see figure S8 in Supplementary Data), the biggest root of (2.2.17) in the interval $[0; 1]$ is in the interval $\left[\frac{1}{3}; 1\right]$ which allow to easily compute $\mathcal{B} \cdot (w_j^1)_{LS}$ and have a straightforward solution between 0 and $\mathcal{B} \cdot (w_j^1)_{LS}$.

Finally, **in the case of L$_1$ penalization**, considering one of the optimization step, e.g. (2.2.11) of ALS optimization, the solution turns out to be an element-wise soft-thresholding of the least square solution $(w_j^1)_{LS}$ $j = 1, \ldots, I_1$ leading to:

$$\left(w_j^1\right)_{L_1} = \begin{cases} 0 & \text{, if } j \in \mathcal{L}_1 \text{ and } \left(w_j^1\right)_{LS} \leq ThresholdL_1 \\ sign\left(\left(w_j^1\right)_{LS}\right)\left(\left|\left(w_j^1\right)_{LS}\right| - ThresholdL_1\right) & \text{, if } i \in \mathcal{L}_1 \text{ and } \left(w_j^1\right)_{LS} > ThresholdL_1, \\ \left(w_j^1\right)_{LS} & \text{otherwise} \end{cases}$$

$$ThresholdL_1 = \frac{\lambda_1}{\|\mathbf{w}^3 \otimes \mathbf{w}^2\|^2}.$$

### 2.2.2. Penalized PARAFAC in the PREW-NPLS algorithm

Penalized PARAFAC based tensor decomposition is integrated into REW-NPLS algorithm to extract iteratively the set of penalized projectors $\left\{\mathbf{w}_f^1 \in \mathbb{R}^{I_1}, \mathbf{w}_f^2 \in \mathbb{R}^{I_2}, \mathbf{w}_f^3 \in \mathbb{R}^{I_3}\right\}_{f=1}^F$ from $\underline{XY}_u$ for each latent space dimension $f \subset \{1, 2, \ldots, F\}$.

For $f = 1$, all the projector elements can be potentially penalized. Therefore the protection set is initialized to $\mathcal{L}_{i,1} \subset \{1, 2, \ldots, I_j\}$ as each projector elements can be penalized. For any $f$, after that the PARAFAC convergence criteria are reached, indices of non-zero elements of $\mathbf{w}_f^i$ (non-penalized projector elements) are removed from $\mathcal{L}_{i,f}$ resulting in the protection set for the next iteration $\mathcal{L}_{i,f+1}$. The protection variable is introduced because REW-NPLS model is estimated via an incremental procedure, the model at iteration $f + 1$ contains information extracted at iteration $f$. Therefore, if a decomposition factor has a non-zero value at iteration $f$, it must be considered at iteration $f + 1$. A scheme representing the basic steps of the PREW-NPLS main loop for a specific $f$ is represented in the case of spatial $L_1$ penalization with a penalization factor $\lambda$ in the Supplementary Materials (Figure S1) as well as a loop of the penalized PARAFAC during this specific PREW-NPLS algorithm. With the exception of the penalized PARAFAC decomposition, PREW-NPLS model calibration is similar than REW-NPLS algorithms. At each iteration $u$, a set of $F$ models is evaluated with a penalization factor $\lambda$. and is noted $\theta_{u,\lambda} = \left\{\underline{\mathbf{Beta}}_u^{f,\lambda}, \underline{\mathbf{bias}}_u^{f,\lambda}\right\}_{f=1}^F$.

The previously presented PREW-NPLS based on the regularized PARAFAC procedure allows to perform group-wise parameter penalization for a fixed penalization hyperparameter $\lambda$. The selection of this hyperparameter influences greatly the sparsity of the solution and the global performance of the algorithm. The selection of the $\lambda$ hyperparameter may be a complex task and is often optimized based on random or grid search using cross-validation strategy. However such strategy cannot be applied for online decoding because they require high computing resources, too long computing time and are not suited to data-flow processing. Therefore, during online experiments, penalization factor $\lambda$ is fixed using prior knowledge or preliminary offline studies.

### 2.3. Experiments

This study rely on neural signals dataset during experiments a patient at CLINATEC®. The « BCI and Tetraplegia » clinical trial was catalogued the 11/09/2015 in the publically accessible register named ClinicalTrials.gov, under the identifier: NCT02550522 [87,88]. The clinical trial was approved by the French authorities: National Agency for the Safety of Medicines and Health Products (Agence nationale de sécurité du médicament et des produits de santé: ANSM) with the registration Number: 2015-A00650-49 and the ethical Committee for the Protection of Individuals (Comité de Protection des Personnes - CPP) with the Registration number: 15-CHUG-19. All research activities were carried out in accordance with the guidelines and regulations of the ANSM and the CPP. The patient signed informed consent for the clinical trial, publication as well as consent to publish the information/image(s) in an online open access publication. Details of the clinical trial protocol are available in [89].

Before to be evaluated during real-time closed loop experiments, PREW-NPLS performance evaluation was achieved during offline studies with pseudo-online procedure. Pseudo-online experiments are offline simulations conducted using the same parameters as those used for the online experiments. Pre-processing, buffer size, batch-wise training and application of the model are performed following the same procedure as that used for online real-time experiments to reproduce the online experiment conditions. Pseudo-online comparison is not fully generalizable for the online case. Nevertheless, it allows characterising the studied algorithms before to perform the online experiments. The datasets used for the pseudo-online comparison were recorded during online closed-loop experiments presented in previous article [89].

The patient performed real-time asynchronous closed-loop 8D experiments using the REW-MSLM incremental adaptive closed-loop decoder. The session clustered 3D alternative two-handed reaching tasks ($AS_{LH}$ and $AS_{RH}$), 1D wrist rotation movements for each hand ($AS_{LW}$ and $AS_{RW}$) and idle state (IS) for a total of $z = 5$ states and 8 continuous dimensions [89]. The number of experts was fixed to N=2 with one expert associated to left limb decoding whereas the other estimated the right limb model. The hand and wrist continuous movements from the same body side were decoded in the same expert. These 8D experiment paradigm was achieved using a virtual avatar effector as visual.

During a session, the patient aimed to reach the proposed targets or rotate the wrist to specific angle. 22 targets were 3D symmetrically distributed in two cubes in front of the patient. The average sessions duration was $29 \pm 8$ min. Six closed-loop experiments were achieved in late September for incremental real-time model adaptation. The total training time of the models for virtual avatar was 3 hours and 37 minutes with a total of 189, 194, 181 and 218 trials for the left and right hand translation and left and right hand rotation control, respectively. The performance of the models were evaluated during 37 avatar experiments distributed over 5 to 203 days after the last model recalibration session (468 to 666 days after implantation). Only the left and right hand 3D movements and related neural signals were used for pseudo-online evaluation of the PREW-NPLS algorithm.

### 2.4. Evaluated models

$L_p$-Penalized REW-NPLS algorithm is evaluated for p=0, 0.5 and 1 and compared to their non-penalized version the REW-NPLS algorithm [21]. For each penalization type, several models with different penalization hyperparameter $\lambda$ are estimated. At a specific penalization hyperparameter $\lambda$, the penalization is maximal and higher penalization hyperparameter lead to exactly equivalent models. Therefore, for each penalization type, several models are computed with increasing penalization hyperparameter $\lambda$ starting at $\lambda = 0$ (the REW-NPLS algorithm) and until the penalization threshold is reached. At the end, 25, 21 and 21 models were evaluated (considering REW-NPLS model) for $L_0$, $L_{0.5}$ and $L_1$-Penalized REW-NPLS algorithm respectively.

### 2.5. Performance indicators

The predicted trajectories performed during the online closed-loop experiments are related to the decoding model currently used during the experiments and patient's feedback. Therefore, usual trajectory decoding indicators cannot be use to evaluate the performance of different algorithms in pseudo-online experiments. A sample-based indicators is introduced to compare the continuous predictions of several algorithms. The dot product indicator $DotP$, known in other field as the cosine similarity, is based on the comparison between the predicted directions $\hat{\mathbf{y}}_t$ and the optimal prediction defined as the 3D Cartesian vector between the current position and the target $\mathbf{y}_t$ for 3D translation tasks using the scalar product. After normalization:

$$DotP = \frac{1}{T}\sum_{t=1}^{T}\frac{\mathbf{y}_t \cdot \hat{\mathbf{y}}_t}{\|\mathbf{y}_t\|\|\hat{\mathbf{y}}_t\|},$$

where "·" defined the dot product, $DotP \in [-1,1]$, $T$ is the number of samples recorded for a specific limb (right or left hand). The average dot product over time provides an indicator of the algorithm's global static prediction. To our knowledge, this indicator was only referenced in three articles where EEG neural signals are analysed [90–92]. This indicator was often used in the information retrieval, text mining and data mining fields [93–95]. The median, 95% confidence interval of the median, 25th and 75th percentiles of the $DotP(t)$ are estimated for each model.

The proposed algorithms aims to converge into sparse solutions by fixing non-relevant (non-informative / noisy) electrodes to exactly 0. Direct decoding performance is therefore not the only relevant indicator. A sparse decoder with the same performance than a "classic" decoder may lead to faster model application and better generalization of the decoded neural signals.

Considering a penalized model $\theta_{u,\lambda_i} = \left\{\underline{\mathbf{Beta}}_u^{f,\lambda_i}, \underline{\mathbf{bias}}_u^{f,\lambda_i}\right\}_{f=1}^{F}$ with $\underline{\mathbf{Beta}}_u^{f,\lambda_i} \in \mathbb{R}^{(I_1 \times ... \times I_M) \times (J_1 \times ... \times J_N)}$ which was estimated using PREW-NPLS with the group-wise penalization restricted to the dimension $I_m$. This model was computed from the set of penalized projectors $\left\{\mathbf{w}_f^1 \in \mathbb{R}^{I_1}, ..., \mathbf{w}_f^M \in \mathbb{R}^{I_M}\right\}_{f=1}^{F}$ evaluated with the penalized PARAFAC decomposition. The model sparsity is defined by the number of element $w_{j,f}^m$ of $\mathbf{w}_f^m \in \mathbb{R}^{I_m}$ fixed to zero. The $SparseIdx$ of the model $\Theta(f, \lambda_n)$ following the dimension $I_m$ is defined as:

$$SparseIdx(\Theta(f, \lambda_n), m) = \frac{\sum_{j=1}^{I_m} \delta_{w_{j,f,j}^m}}{I_m}.$$

Here, $\delta$ is the Kronecker symbol.

## 3. Results

The Dot product performance and the sparsity index of the $L_0$, $L_{0.5}$ and $L_1$ models for the left and the right hand movement tasks are presented depending on the penalization coefficient $\lambda$ in Figure 1. The results are presented using the median, the $25^{th}$ ($Q_1$) and $75^{th}$ ($Q_3$) percentiles using the notation: median ($Q_1 - Q_3$).

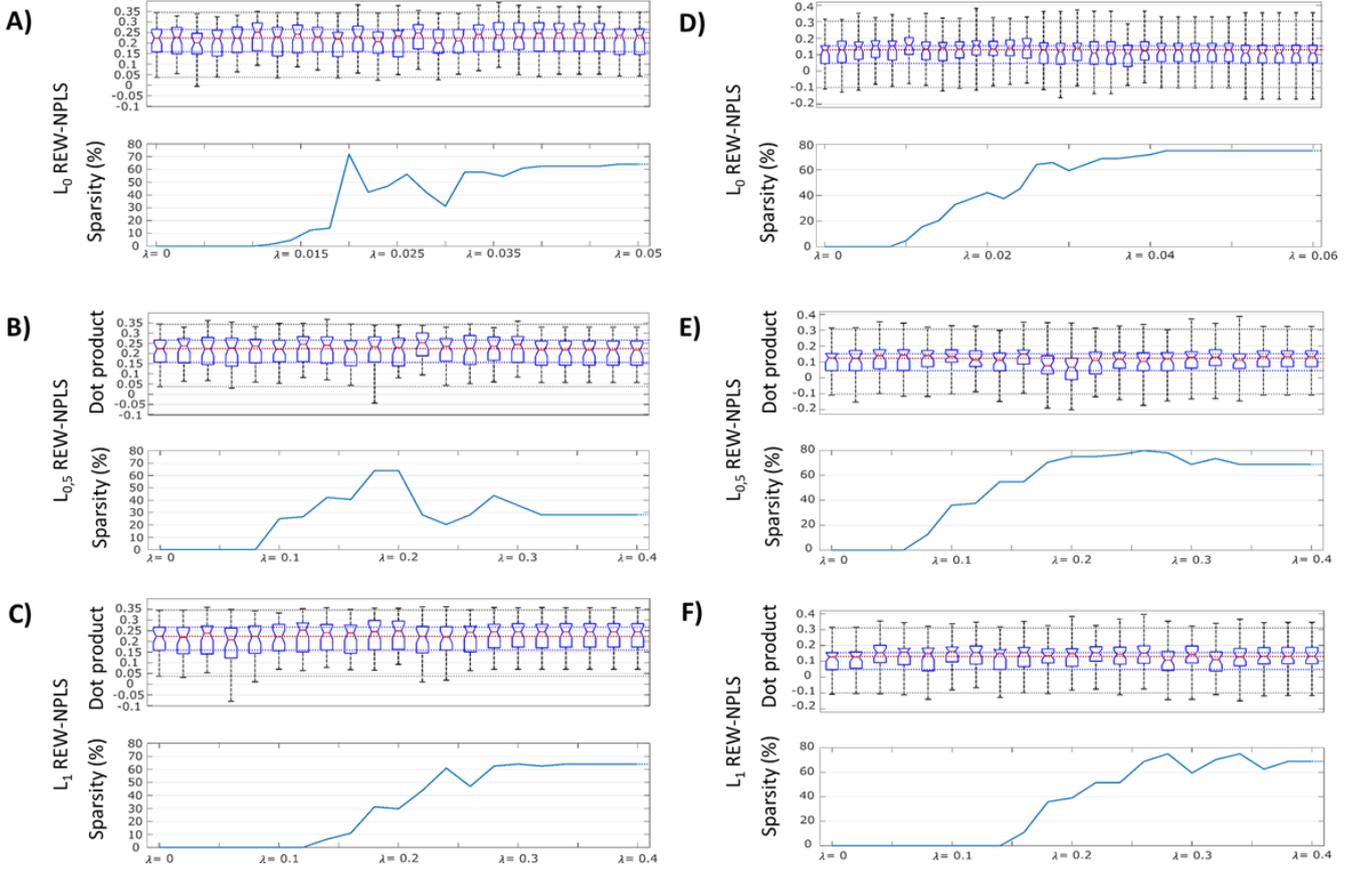

*Figure 1 : The model performance indicators of the $L_p$ REW-NPLS algorithm for left and right hand movement decoding. The cosine similarity and the model sparsity were computed for the $L_0$ REW-NPLS (A), $L_{0.5}$ REW-NPLS (B) and $L_1$ REW-NPLS (C) models estimated for the left hand decoding. Similarly, $L_0$ (D), $L_{0.5}$ (E) and $L_1$ (F) REW-NPLS model performances computed for the right hand 3D movement decoding are presented. The cosine similarly performance on each session was summarized using a box plot representation where the red line is the median the blue lines indicate the $25^{th}$ and $75^{th}$ percentiles ($Q_1$ and $Q_3$). Additionally, the black boundaries show the upper and lower extreme cosine similarity obtained for the experiments. The performance of the REW-NPLS algorithms is presented in the first box plot of each $L_p$ REW-NPLS algorithm sub-plot. The median, $Q_1$ and $Q_3$ of the REW-NPLS models are extended using horizontal dotted lines for easier performance comparison. Additionally, the sparsity of each solution is depicted.*

### 3.1. Left hand decoding performance

**The state of the art REW-NPLS ($\lambda = 0$)** performance, presented in the first position of each sub figures (Figure 1A, B and C), highlighted a median of 0.22, a $Q_1 = 0.158$ and a $Q_3 = 0.266$ which is noted $0.223$ ($0.158 - 0.266$) for the left hand decoding study.

**$L_0$ REW-NPLS algorithms** (Figure 1A) show relevant performance for different penalization coefficient $\lambda$ value. Obviously the $SparseIdx$ indicator increases with higher penalization coefficient value. However, the the dot product (cosine similarity) highlighted better performance than REW-NPLS algorithm with various $\lambda$ value. For $\lambda = 0.01$, the $SparseIdx = 0\%$ but the dot product (cosine similarity) was evaluated at $0.252\ (0.165 - 0.296)$. For $\lambda = 0.026$, the cosine similarity was estimated at $0.248\ (0.173 - 0.288)$ with a $SparseIdx = 56,25\%$. For $\lambda = 0.04$ to $\lambda = 0.046$ and for $\lambda > 0.046$ PREW-NPLS demonstrated a $0.248\ (0.162 - 0.294)$ and $0.236\ (0.146 - 0.268)$ continuous decoding performance with 40 ($SparseIdx = 62.5\%$) and 41 ($SparseIdx = 64.06\%$) electrode parameters over 64 set to 0 value.

**$L_{0.5}$ REW-NPLS algorithm** (Figure 1B) does not present decoding performance improvements compared to REW-NPLS algorithm with the exception of some models. For $\lambda = 0.22$ with 18 electrodes parameter weights set to zero ($SparseIdx = 28.13\%$) the PREW-NPLS model highlighted higher cosine similarity performance $0.253\ (0.189 - 0.301)$ than the REW-NPLS model. Additionally, for $\lambda = 0.3$, the dot product was estimated at $0.245\ (0.156 - 0.2838)$ with a sparsity index $SparseIdx = 35.94\%$. Finally, for $\lambda > 0.32$ the model sparsity converged to $SparseIdx = 28.13\%$ with continuous decoding performance similar to REW-NPLS model: $0.217\ (0.143 - 0.261)$.

**$L_1$ REW-NPLS algorithm** (Figure 1C) highlighted similar results than $L_0$ and $L_{0.5}$ REW-NPLS algorithms. For $\lambda = 0.12$, with a sparsity of $SparseIdx = 0\%$, the PREW-NPLS model highlighted a $0.253\ (0.151 - 0.286)$ cosine similarity. A sparsity index of $SparseIdx = 29.69\%$ is reached for $\lambda = 0.20$ with a decoding performance of $0.249\ (0.162 - 0.295)$. Finally, for $\lambda > 0.34$, 41 electrodes parameter weights are set to 0 value leading to a $0.245\ (0.173 - 0.283)$.

The REW-NPLS and the $L_p$ REW-NPLS model parameter weights are illustrated on the temporal, frequency and spatial domain in the Supplementary Materials Figure S2 for the left hand models. For easier comparison and selection the presented models are the ones with "converged" penalization coefficient λ=0.06,0.4 and ,0.4 for $L_0$, $L_{0.5}$ and $L_1$ REW-NPLS algorithms respectively. Moreover, the spatial parameter weights are presented in the Supplementary Materials Figure S4 for the left hand models on a map with the electrode locations relative to the sensory (SS) and motor (MS) sulci.

### 3.2. Right hand decoding performance

A similar study has been performed to decode the right hand movements. No statistical difference in the cosine similarity indicator was highlighted between the state of the art REW-NPLS algorithm and the $L_0$, $L_{0.5}$ and $L_1$ REW-NPLS models. The decoding performance of all the models for all penalization parameter were highly variable between sessions (high inter-session variability). Right arm decoding study stresses worse cosine similarity than left arm decoding study.

**The state of the art REW-NPLS ($\lambda = 0$) model,** presented for the right arm translation decoding in the first position of Figure 1D, E and F, highlighted a cosine similarity of $0.127\ (0.0468 - 0.155)$.

**$L_0$ REW-NPLS algorithms** (Figure 1D) show performance improvements with sparse solutions for different penalization coefficient $\lambda$ value. For $\lambda = 0.01$, the $SparseIdx = 4.68\%$ corresponding to only 3 electrode parameter weights set to zero value but the dot product (cosine similarity) was evaluated at $0.157\ (0.1018 - 0.203$. These performance represent a cosine similarity enhancement of 24%, 117% and 30% for the median, the $Q_1$ and $Q_3$ metrics respectively. For $\lambda = 0.018$, the cosine similarity was estimated at $0.157\ (0.0989 - 0.185)$ with a $SparseIdx = 37.5\%$. For $\lambda = 0.024$, sparser solutions are obtained with a sparsity index of $SparseIdx = 45.31\%$ with a cosine similarity estimated at $0.153\ (0.0786 - 0.198)$. For $\lambda > 0.04$ $L_0$ PREW-NPLS models converge to very sparse solution with 48 electrodes over 64 ($SparseIdx = 75\%$) removed from the solution which highlighted decoding performance similar to the state of the art REW-NPLS. The best performance of the models with $\lambda > 0.04$ was estimated at $0.128\ (0.058 - 0.168)$.

**$L_{0.5}$ REW-NPLS algorithm** (Figure 1E), similarly than $L_0$ REW-NPLS decoder highlighted better decoding performance than REW-NPLS with sparser solutions for some penalization parameter $\lambda$. For $\lambda = 0.1$, 23 electrodes were removed from the model ($SparseIdx = 35.94\%$) and the cosined similarity was estimated at $0.136\ (0.100 - 0.177)$. With higher penalization parameter $\lambda = 0.16$ sparser model is computed with $SparseIdx = 54.69\%$ without decreasing the decoding performance $DotP = 0.150\ (0.0881 - 0.176)$. The sparsest models are obtained for $\lambda = 0.26$ and $\lambda = 0.28$ showing a sparsity index of $SparseIdx = 79.69\%$ and $SparseIdx = 78.13\%$ respectively. Finally, the models converged to the same solutions for $\lambda > 0.36$ with a sparsity of $SparseIdx =$ 68.75% (44 electrodes removed from the final solution) and a cosine similarity of $DotP = 0.131\ (0.0835 - 0.186)$.

**$L_1$ REW-NPLS algorithm** (Figure 1F) results show better (not significantly) decoding performance than REW-NPLS algorithm for numerous penalization parameter $\lambda$. Several models with small penalization parameter $\lambda = 0.04, 0.06$ and $0.1$ without setting any electrode to zero ($SparseIdx = 0\%$) highlighted a cosine similarity of $DotP = 0.154\ (0.0915 - 0.202)$, $DotP = 0.158\ (0.0791 - 0.184)$ and $DotP = 0.164\ (0.0959 - 0.191)$ representing a median improvements of 21%, 24% and 29% respectively. Similar decoding performance were obtained for higher penalization parameter $\lambda = 0.22$ and $\lambda = 0.26$ with a dot product indicator of $DotP = 0.154\ (0.101 - 0.192)$ and $DotP = 0.152\ (0.0872 - 0.197)$ but with 33 (51.56%) and 44 (68.75%) electrodes parameters weights set to zero values respectively. Finally, for a penalization parameter $\lambda > 0.38$, the models stabilized to a solution with a sparsity indicator of $SparseIdx = 68.75\%$ with a decoding performance of $DotP = 0.131\ (0.0835 - 0.186)$.

The REW-NPLS and the $L_p$ REW-NPLS model parameter weights are illustrated on the temporal, frequency and spatial domain in the Supplementary Materials Figure S3 for the right hand models. For easier comparison and selection the presented models are the ones with "converged" penalization coefficient λ=0.06, 0.4 and 0.4 for $L_0$, $L_{0.5}$ and $L_1$ REW-NPLS algorithms respectively. Moreover, the spatial parameter weights are presented in the Supplementary Materials Figure S5 for the right hand models on a map with the electrode locations relative to the sensory (SS) and motor (MS) sulci.

## 4. Discussion

The study was focused on pseudo-online decoding of the left or right hand movements recorded during online closed-loop experiments where the patients controlled a virtual avatar effector. The dataset D is composed of 43 experiments. The tested models were calibrated during the offline study using the first 6 experiments (recorded in late September 2018) and was tested based on the experiments recorded between early October 2018 and mid-March 2019. The number of training session was small (14%) and focused at the beginning of the experiments (no re-calibration period). Therefore, high inter-session variability is observed between the best and the worst PREW-NPLS decoding performance. Consequently, it is complex to extract statistical evidence of the PREW-NPLS decoding superiority. However, the PREW-NPLS algorithms highlighted equivalent or better decoding performance than REW-NPLS decoder using sparse solutions with up to 80% of the electrodes set to 0 value for the right hand models. Additionally, $L_0$, $L_{0.5}$ and $L_1$ REW-NPLS algorithms converged to similar solutions with comparable decoding performance. However, the $L_{0.5}$ REW-NPLS algorithm is looking for an exact solution of a cubic polynomial function which requires an important computational burden to be solved and may not be adapted to online CLDA. All the PREW-NPLS required to fixed the penalization hyperparameter $\lambda$ before to estimate the model. The PREW-NPLS did not highlighted higher decoding performance and sparse solution for every penalization hyperparameter. Therefore, it is required to optimize the penalization hyperparameter $\lambda$ before to integrate the solution into online closed-loop decoding experiments. Numerous models were created offline to find the best penalization hyperparameter. This offline study was highly time-consuming. Furthermore, it did not allow to conclude on an optimal penalization hyperparameter for online closed-loop experiments as the penalization hyperparameter is also dependent of the latent space dimension hyperparameter, may be different during closed-loop experiments due to the neural signals related to the patient's feedback and may change over time.

In order to bypass these limitations the automatic Penalized REW-NPLS (APREW-NPLS) was designed. APREW-NPLS performs an incremental model calibration procedure for a set of models with different penalization hyperparameters in the same time to evaluate the best model across time.

## Acknowledgements

CLINATEC® is a Laboratory of CEA-Grenoble and has statutory links with the University Hospital of Grenoble (CHUGA) and with University Grenoble Alpes (UGA). This study was funded by CEA (recurrent funding) and the French Ministry of Health (Grant PHRC-15-15-0124), Institut Carnot, Fonds de Dotation CLINATEC®. Fondation Philanthropique Edmond J Safra is a major founding institution of the CLINATEC® Edmond J Safra Biomedical Research Center.

We thank Thomas Costecalde, Benoit Milville, Serpil Karakas, Guillaume Charvet, Jean-Claude Royer, Stéphane Pezzani, Stéphan Chabardes and the CHUGA members working at CLINATEC® for the help in the experimental setup and clinical trial experiments.

## Declaration of interests

We declare no competing interests.

## References

[1] Nicolas-Alonso L F and Gomez-Gil J 2012 Brain computer interfaces, a review *Sensors* **12** 1211–79

[2] Bellman R E 1961 *Adaptive Control Processes: A Guided Tour* (Princeton University Press)

[3] Bishop C M 2006 *Pattern Recognition and Machine Learning* (Springer New York)

[4] Remeseiro B and Bolon-Canedo V 2019 A review of feature selection methods in medical applications *Comput. Biol. Med.* **112** 103375

[5] Brandman D M, Hosman T, Saab J, Burkhart M C, Shanahan B E, Ciancibello J G, Sarma A A, Milstein D J, Vargas-Irwin C E, Franco B, Jessica Kelemen, Blabe C, Murphy B A, Young D R, Willett F R, Pandarinath C, Stavisky S D, Kirsch R F, Walter B L, Ajiboye A B, Cash S S, Eskandar E N, Miller J P, Sweet J A, Shenoy K V, Henderson J M, Jarosiewicz B, Harrison M T, Simeral J D and Hochberg L R 2018 Rapid calibration of an intracortical brain–computer interface for people with tetraplegia *J. Neural Eng.* **15** 026007

[6] Haufe S, Dähne S and Nikulin V V 2014 Dimensionality reduction for the analysis of brain oscillations *NeuroImage* **101** 583–97

[7] Lotte F, Bougrain L, Cichocki A, Clerc M, Congedo M, Rakotomamonjy A and Yger F 2018 A review of classification algorithms for EEG-based brain–computer interfaces: a 10 year update *J. Neural Eng.* **15** 031005

[8] Boussetta R, El Ouakouak I, Gharbi M and Regragui F 2018 EEG Based Brain Computer Interface for Controlling a Robot Arm Movement Through Thought *IRBM* **39** 129–35

[9] Sreenath R and Ramana R 2017 Classification of denoising techniques for EEG signals: A review *Int. J. Pure Appl. Math.* **117** 967–72

[10] Schaeffer M-C and Aksenova T 2016 Switching Markov decoders for asynchronous trajectory reconstruction from ECoG signals in monkeys for BCI applications *J. Physiol.-Paris* **110** 348–60

[11] Khan J, Bhatti M H, Khan U G and Iqbal R 2019 Multiclass EEG motor-imagery classification with sub-band common spatial patterns *EURASIP J. Wirel. Commun. Netw.* **2019** 174

[12] Jiang T, Jiang T, Wang T, Mei S, Liu Q, Li Y, Wang X, Prabhu S, Sha Z and Ince N F 2017 Characterization and Decoding the Spatial Patterns of Hand Extension/Flexion using High-Density ECoG *IEEE Trans. Neural Syst. Rehabil. Eng.* **25** 370–9

[13] Seifzadeh S, Rezaei M, Faez K and Amiri M 2017 Fast and Efficient Four-class Motor Imagery Electroencephalography Signal Analysis Using Common Spatial Pattern-Ridge Regression Algorithm for the Purpose of Brain-Computer Interface *J. Med. Signals Sens.* **7** 80–5

[14] Marathe A R and Taylor D M 2013 Decoding continuous limb movements from high-density epidural electrode arrays using custom spatial filters *J. Neural Eng.* **10** 036015

[15] Sannelli C, Vidaurre C, Müller K-R and Blankertz B 2016 Ensembles of adaptive spatial filters increase BCI performance: an online evaluation *J. Neural Eng.* **13** 046003

[16] Jafarifarmand A and Badamchizadeh M A 2020 Real-time multiclass motor imagery brain-computer interface by modified common spatial patterns and adaptive neuro-fuzzy classifier *Biomed. Signal Process. Control* **57** 101749

[17] Hsu S H, Mullen T R, Jung T P and Cauwenberghs G 2016 Real-Time Adaptive EEG Source Separation Using Online Recursive Independent Component Analysis *IEEE Trans. Neural Syst. Rehabil. Eng.* **24** 309–19

[18] Palmer J A and Hirata M 2018 Independent Component Analysis (ICA) Features for Electro-corticographic (ECoG) Brain-Machine Interfaces (BMIs) 臨床神経生理学 **46** 55–60

[19] Bundy D T, Pahwa M, Szrama N and Leuthardt E C 2016 Decoding three-dimensional reaching movements using electrocorticographic signals in humans *J. Neural Eng.* **13** 026021

[20] Choi H, Lee J, Park J, Lee S, Ahn K, Kim I Y, Lee K-M and Jang D P 2018 Improved prediction of bimanual movements by a two-staged (effector-then-trajectory) decoder with epidural ECoG in nonhuman primates *J. Neural Eng.* **15** 016011

[21] Eliseyev A, Auboiroux V, Costecalde T, Langar L, Charvet G, Mestais C, Aksenova T and Benabid A-L 2017 Recursive Exponentially Weighted N-way Partial Least Squares Regression with Recursive-Validation of Hyper-Parameters in Brain-Computer Interface Applications *Sci. Rep.* **7** 16281

[22] Kim S-P, Sanchez J C, Rao Y N, Erdogmus D, Carmena J M, Lebedev M A, Nicolelis M a. L and Principe J C 2006 A comparison of optimal MIMO linear and nonlinear models for brain-machine interfaces *J. Neural Eng.* **3** 145–61

[23] Khaire U M and Dhanalakshmi R 2019 Stability of feature selection algorithm: A review *J. King Saud Univ. - Comput. Inf. Sci.*

[24] Bolón-Canedo V, Sánchez-Maroño N and Alonso-Betanzos A 2013 A review of feature selection methods on synthetic data *Knowl. Inf. Syst.* **34** 483–519

[25] Hastie T, Tibshirani R, Wainwright M, Tibshirani R and Wainwright M 2015 *Statistical Learning with Sparsity : The Lasso and Generalizations* (Chapman and Hall/CRC)

[26] Sreeja S R, Himanshu, Samanta D and Sarma M 2019 Weighted sparse representation for classification of motor imagery EEG signals *2019 41st Annual International Conference of the IEEE Engineering in Medicine and Biology Society (EMBC)* 2019 41st Annual International Conference of the IEEE Engineering in Medicine and Biology Society (EMBC) pp 6180–3

[27] Flamary R and Rakotomamonjy A 2012 Decoding Finger Movements from ECoG Signals Using Switching Linear Models *Front. Neurosci.* **6**

[28] López-Larraz E, Montesano L, Gil-Agudo Á and Minguez J 2014 Continuous decoding of movement intention of upper limb self-initiated analytic movements from pre-movement EEG correlates *J. NeuroEngineering Rehabil.* **11** 153

[29] Lotte F and Guan C 2011 Regularizing Common Spatial Patterns to Improve BCI Designs: Unified Theory and New Algorithms *IEEE Trans. Biomed. Eng.* **58** 355–62

[30] Eliseyev A, Moro C, Faber J, Wyss A, Torres N, Mestais C, Benabid A L and Aksenova T 2012 L1-Penalized N-way PLS for subset of electrodes selection in BCI experiments *J. Neural Eng.* **9** 045010

[31] Zhang Y, Zhou G, Jin J, Wang M, Wang X and Cichocki A 2013 L1-Regularized Multiway Canonical Correlation Analysis for SSVEP-Based BCI *IEEE Trans. Neural Syst. Rehabil. Eng.* **21** 887–96

[32] Cincotti F, Mattia D, Aloise F, Bufalari S, Astolfi L, De Vico Fallani F, Tocci A, Bianchi L, Marciani M G, Gao S, Millan J and Babiloni F 2008 High-resolution EEG techniques for brain–computer interface applications *J. Neurosci. Methods* **167** 31–42

[33] Nagel S and Spüler M 2019 Asynchronous non-invasive high-speed BCI speller with robust non-control state detection *Sci. Rep.* **9** 1–9

[34] Kim S, White A, Scalzo F and Collier D 2018 Elastic net ensemble classifier for event-related potential based automatic spelling *Biomed. Signal Process. Control* **46** 166–73

[35] Peterson V, Wyser D, Lambercy O, Spies R and Gassert R 2019 A penalized time-frequency band feature selection and classification procedure for improved motor intention decoding in multichannel EEG *J. Neural Eng.* **16** 016019

[36] Eliseyev A and Aksenova T 2016 Penalized Multi-Way Partial Least Squares for Smooth Trajectory Decoding from

Electrocorticographic (ECoG) Recording ed D Zhang *PLOS ONE* **11** e0154878

[37] Nakanishi Y, Yanagisawa T, Shin D, Kambara H, Yoshimura N, Tanaka M, Fukuma R, Kishima H, Hirata M and Koike Y 2017 Mapping ECoG channel contributions to trajectory and muscle activity prediction in human sensorimotor cortex *Sci. Rep.* **7** 1–13

[38] Toda A, Imamizu H, Kawato M and Sato M 2011 Reconstruction of two-dimensional movement trajectories from selected magnetoencephalography cortical currents by combined sparse Bayesian methods *NeuroImage* **54** 892–905

[39] Mishra P K, Jagadish B, Kiran M P R S, Rajalakshmi P and Reddy D S 2018 A Novel Classification for EEG Based Four Class Motor Imagery Using Kullback-Leibler Regularized Riemannian Manifold *2018 IEEE 20th International Conference on e-Health Networking, Applications and Services (Healthcom)* 2018 IEEE 20th International Conference on e-Health Networking, Applications and Services (Healthcom) pp 1–5

[40] Giordani P and Rocci R 2013 Constrained Candecomp/Parafac via the Lasso *Psychometrika* **78** 669–84

[41] Martínez-Montes E, Sánchez-Bornot J M and Valdés-Sosa P A 2008 PENALIZED PARAFAC ANALYSIS OF SPONTANEOUS EEG RECORDINGS *Stat. Sin.* **18** 1449–64

[42] van Gerven M, Hesse C, Jensen O and Heskes T 2009 Interpreting single trial data using groupwise regularisation *NeuroImage* **46** 665–76

[43] Motrenko A and Strijov V 2018 Multi-way feature selection for ECoG-based Brain-Computer Interface *Expert Syst. Appl.* **114** 402–13

[44] Wu Q, Zhang Y, Liu J, Sun J, Cichocki A and Gao F 2019 Regularized Group Sparse Discriminant Analysis for P300-Based Brain–Computer Interface *Int. J. Neural Syst.* **29** 1950002

[45] Hervás D, Prats-Montalbán J M, García-Cañaveras J C, Lahoz A and Ferrer A 2019 Sparse N-way partial least squares by L1-penalization *Chemom. Intell. Lab. Syst.* **185** 85–91

[46] Kim H J, Ollila E, Koivunen V and Croux C 2013 Robust and sparse estimation of tensor decompositions *2013 IEEE Global Conference on Signal and Information Processing* 2013 IEEE Global Conference on Signal and Information Processing pp 965–8

[47] Kim H J, Ollila E, Koivunen V and Poor H V 2014 Robust iteratively reweighted Lasso for sparse tensor factorizations *2014 IEEE Workshop on Statistical Signal Processing (SSP)* 2014 IEEE Workshop on Statistical Signal Processing (SSP) pp 420–3

[48] Oliver G, Sunehag P and Gedeon T 2013 Online feature selection for Brain Computer Interfaces *2013 IEEE Symposium on Computational Intelligence, Cognitive Algorithms, Mind, and Brain (CCMB)* 2013 IEEE Symposium on Computational Intelligence, Cognitive Algorithms, Mind, and Brain (CCMB) pp 122–9

[49] Robinson N, Guan C, Vinod A P, Ang K K and Tee K P 2013 Multi-class EEG classification of voluntary hand movement directions *J. Neural Eng.* **10** 056018

[50] Kumar S, Sharma A and Tsunoda T 2017 An improved discriminative filter bank selection approach for motor imagery EEG signal classification using mutual information *BMC Bioinformatics* **18** 545

[51] Schroder M, Bogdan M, Hinterberger T and Birbaumer N 2003 Automated EEG feature selection for brain computer interfaces *First International IEEE EMBS Conference on Neural Engineering, 2003. Conference Proceedings.* First International IEEE EMBS Conference on Neural Engineering, 2003. Conference Proceedings. pp 626–9

[52] Garrett D, Peterson D A, Anderson C W and Thaut M H 2003 Comparison of linear, nonlinear, and feature selection methods for EEG signal classification *IEEE Trans. Neural Syst. Rehabil. Eng.* **11** 141–4

[53] Brunner C, Scherer R, Graimann B, Supp G and Pfurtscheller G 2006 Online control of a brain-computer interface using phase synchronization *IEEE Trans. Biomed. Eng.* **53** 2501–6

[54] Cantillo-Negrete J, Carino-Escobar R I, Carrillo-Mora P, Elias-Vinas D and Gutierrez-Martinez J 2018 Motor imagery-based brain-computer interface coupled to a robotic hand orthosis aimed for neurorehabilitation of stroke patients *J. Healthc. Eng.* **2018**

[55] Huang D, Lin P, Fei D, Chen X and Bai O 2009 EEG-based online two-dimensional cursor control *2009 Annual International Conference of the IEEE Engineering in Medicine and Biology Society* 2009 Annual International Conference of the IEEE Engineering in Medicine and Biology Society pp 4547–50

[56] Spüler M, Rosenstiel W and Bogdan M 2012 Adaptive SVM-Based Classification Increases Performance of a MEG-Based Brain-Computer Interface (BCI) *Artificial Neural Networks and Machine Learning – ICANN 2012* Lecture Notes in Computer Science ed A E P Villa, W Duch, P Érdi, F Masulli and G Palm (Springer Berlin Heidelberg) pp 669–76

[57] Zhao Q, Liqing Zhang, Cichocki A and Jie Li 2008 Incremental Common Spatial Pattern algorithm for BCI *2008 IEEE International Joint Conference on Neural Networks (IEEE World Congress on Computational Intelligence)* 2008 IEEE International Joint Conference on Neural Networks (IEEE World Congress on Computational Intelligence) pp 2656–9

[58] Chen C K and Fang W C 2017 A reliable brain-computer interface based on SSVEP using online recursive independent component analysis *2017 39th Annual International Conference of the IEEE Engineering in Medicine and Biology Society (EMBC)* 2017 39th Annual International Conference of the IEEE Engineering in Medicine and Biology Society (EMBC) pp 2798–801

[59] Mobaien A and Boostani R 2016 ACSP: Adaptive CSP filter for BCI applications *2016 24th Iranian Conference on Electrical Engineering (ICEE)* 2016 24th Iranian Conference on Electrical Engineering (ICEE) pp 466–71

[60] Ang K K, Chin Z Y, Zhang H and Guan C 2011 Filter Bank Common Spatial Pattern (FBCSP) algorithm using online adaptive and semi-supervised learning *The 2011 International Joint Conference on Neural Networks* The 2011 International Joint Conference on Neural Networks pp 392–6

[61] Song X and Yoon S-C 2015 Improving brain–computer interface classification using adaptive common spatial patterns *Comput. Biol. Med.* **61** 150–60

[62] Woehrle H, Krell M M, Straube S, Kim S K, Kirchner E A and Kirchner F 2015 An Adaptive Spatial Filter for User-Independent Single Trial Detection of Event-Related Potentials *IEEE Trans. Biomed. Eng.* **62** 1696–705

[63] Dagher I 2010 Incremental PCA-LDA algorithm *2010 IEEE International Conference on Computational Intelligence for Measurement Systems and Applications* 2010 IEEE International Conference on Computational Intelligence for Measurement Systems and Applications pp 97–101

[64] Faller J, Vidaurre C, Solis-Escalante T, Neuper C and Scherer R 2012 Autocalibration and Recurrent Adaptation: Towards a Plug and Play Online ERD-BCI *IEEE Trans. Neural Syst. Rehabil. Eng.* **20** 313–9

[65] Mend M and Kullmann W H 2012 Human computer interface with online brute force feature selection *Biomed. Eng. Biomed. Tech.* **57** 659–662

[66] Long J, Gu Z, Li Y, Yu T, Li F and Fu M 2011 Semi-supervised joint spatio-temporal feature selection for P300-based BCI speller *Cogn. Neurodyn.* **5** 387

[67] Andreu-Perez J, Cao F, Hagras H and Yang G 2018 A Self-Adaptive Online Brain–Machine Interface of a Humanoid Robot Through a General Type-2 Fuzzy Inference System *IEEE Trans. Fuzzy Syst.* **26** 101–16

[68] Moro R, Berger P and Bielikova M 2017 Towards adaptive brain-computer interfaces: Improving accuracy of detection of event-related potentials *2017 12th International Workshop on Semantic and Social Media Adaptation and Personalization (SMAP)* 2017 12th International Workshop on Semantic and Social Media Adaptation and Personalization (SMAP) pp 34–9

[69] Ma Z, Cheng J and Tao D 2020 Online learning using projections onto shrinkage closed balls for adaptive brain-computer interface *Pattern Recognit.* **97** 107017

[70] Shin Y, Lee S, Ahn M, Cho H, Jun S C and Lee H-N 2015 Noise robustness analysis of sparse representation based classification method for non-stationary EEG signal classification *Biomed. Signal Process. Control* **21** 8–18

[71] Roijendijk L, Gielen S and Farquhar J 2016 Classifying Regularized Sensor Covariance Matrices: An Alternative to CSP *IEEE Trans. Neural Syst. Rehabil. Eng.* **24** 893–900

[72] Sharghian V, Rezaii T Y, Farzamnia A and Tinati M A 2019 Online Dictionary Learning for Sparse Representation-Based Classification of Motor Imagery EEG *2019 27th Iranian Conference on Electrical Engineering (ICEE)* 2019 27th Iranian Conference on Electrical Engineering (ICEE) pp 1793–7

[73] Sheikhattar A, Fritz J B, Shamma S A and Babadi B 2015 Adaptive sparse logistic regression with application to neuronal plasticity analysis *2015 49th Asilomar Conference on Signals, Systems and Computers* 2015 49th Asilomar Conference on Signals, Systems and Computers pp 1551–5

[74] Yang C, Qiao J, Ahmad Z, Nie K and Wang L 2019 Online sequential echo state network with sparse RLS algorithm for time series prediction *Neural Netw.* **118** 32–42

[75] Chen B, Zhao S, Seth S and Principe J C 2012 Online efficient learning with quantized KLMS and L1 regularization *The 2012 International Joint Conference on Neural Networks (IJCNN)* The 2012 International Joint Conference on Neural Networks (IJCNN) pp 1–6

[76] Foodeh R, Ebadollahi S and Daliri M R 2020 Regularized Partial Least Square Regression for Continuous Decoding in Brain-Computer Interfaces *Neuroinformatics*

[77] Cichocki A, Mandic D, Lathauwer L D, Zhou G, Zhao Q, Caiafa C and PHAN H A 2015 Tensor Decompositions for Signal Processing Applications: From two-way to multiway component analysis *IEEE Signal Process. Mag.* **32** 145–63

[78] Kolda T and Bader B 2009 Tensor Decompositions and Applications *SIAM Rev.* **51** 455–500

[79] Pereira Da Silva A, Comon P and De Almeida A L F 2015 *Rank-1 Tensor Approximation Methods and Application to Deflation* (GIPSA-lab)

[80] Tomasi G 2006 *Practical and Computational Aspects in Chemometric Data Analysis* (Denmark: The Royal Veterinary and Agricultural University, Frederiksberg)

[81] Faber N (Klaas) M, Bro R and Hopke P K 2003 Recent developments in CANDECOMP/PARAFAC algorithms: a critical review *Chemom. Intell. Lab. Syst.* **65** 119–37

[82] Tomasi G and Bro R 2006 A comparison of algorithms for fitting the PARAFAC model *Comput. Stat. Data Anal.* **50** 1700–34

[83] Silva A P da, Comon P and Almeida A L F de 2015 An iterative deflation algorithm for exact CP tensor decomposition *2015 IEEE International Conference on Acoustics, Speech and Signal Processing (ICASSP)* 2015 IEEE International Conference on Acoustics, Speech and Signal Processing (ICASSP) pp 3961–5

[84] Chen Bilian, He Simai, Li Zhening and Zhang Shuzhong 2012 Maximum Block Improvement and Polynomial Optimization *SIAM J. Optim.* **22** 87–107

[85] Wang Liqi and Chu M T 2014 On the Global Convergence of the Alternating Least Squares Method for Rank-One Approximation to Generic Tensors *SIAM J. Matrix Anal. Appl.* **35** 1058–72

[86] Uschmajew A 2015 A new convergence proof for the higher-order power method and generalizations *ArXiv14074586 Math*

[87] Anon Brain Computer Interface: Neuroprosthetic Control of a Motorized Exoskeleton ClinicalTrials.gov

[88] Anon ICTRP Search Portal. CLINICAL TRIAL NCT02550522

[89] Benabid A L, Costecalde T, Eliseyev A, Charvet G, Verney A, Karakas S, Foerster M, Lambert A, Morinière B, Abroug N, Schaeffer M-C, Moly A, Sauter-Starace F, Ratel D, Moro C, Torres-Martinez N, Langar L, Oddoux M, Polosan M, Pezzani S, Auboiroux V, Aksenova T, Mestais C and Chabardes S 2019 An exoskeleton controlled by an epidural wireless brain–machine interface in a tetraplegic patient: a proof-of-concept demonstration *Lancet Neurol.* **0**

[90] Olcay B O and Karaçalı B 2019 Evaluation of synchronization measures for capturing the lagged synchronization between EEG channels: A cognitive task recognition approach *Comput. Biol. Med.* **114** 103441

[91] Xu Y, Wei Q, Zhang H, Hu R, Liu J, Hua J and Guo F 2019 Transfer Learning Based on Regularized Common Spatial Patterns Using Cosine Similarities of Spatial Filters for Motor-Imagery BCI *J. Circuits Syst. Comput.*

[92] Rashid U, Niazi I K, Signal N and Taylor D 2018 An EEG Experimental Study Evaluating the Performance of Texas Instruments ADS1299 *Sensors* **18** 3721

[93] Schenker A, Last M, Bunke H and Kandel A 2003 Comparison of Distance Measures for Graph-Based Clustering of Documents *Graph Based Representations in Pattern Recognition* Lecture Notes in Computer Science ed E Hancock and M Vento (Berlin, Heidelberg: Springer) pp 202–13

[94] Rani M S and S S 2017 Perspectives of the performance metrics in lexicon and hybrid based approaches: a review *Int. J. Eng. Technol.* **6** 108–15

[95] Umakanth N and Santhi S 2020 Classification and ranking of trending topics in twitter using tweets text *J. Crit. Rev.* **7** 895–9

# Supplementary Data

The supplementary figures, equations and references included in this section are independent from the core article. All the figures, equations and references indexing are initialized to 1.

## Supplementary Figures

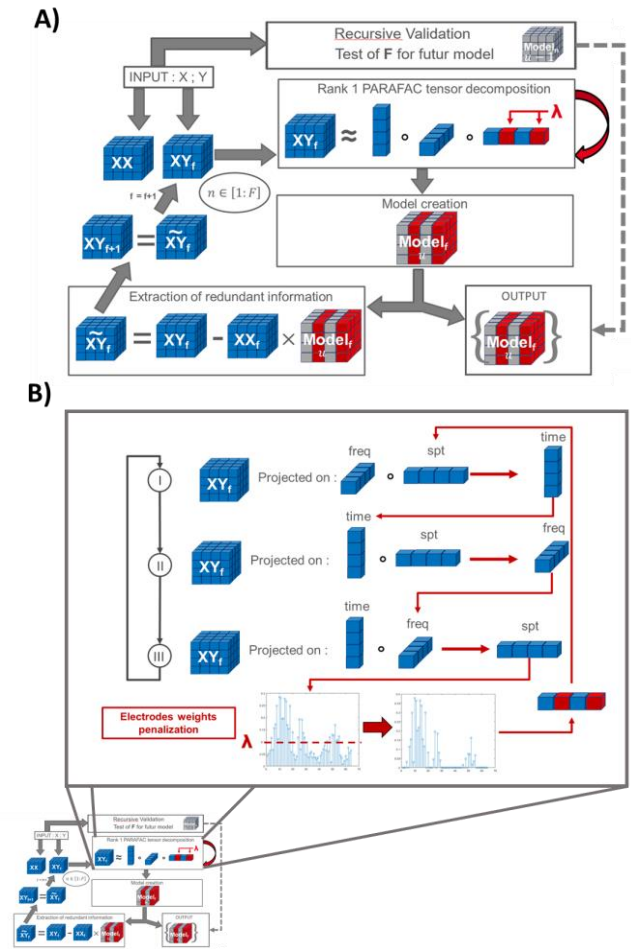

*Figure S1: Penalized REW-NPLS (PREW-NPLS) algorithm. (A) PREW-NPLS algorithm main steps with penalized PARAFAC decomposition leading to slice-wise sparse model. (B) Example of the $L_1$-PARAFAC decomposition performed in the case of $L_1$-PREW-NPLS penalization on the space (electrodes) domain with the hyperparameter λ. ALS algorithm is used for decomposition factor estimation.*

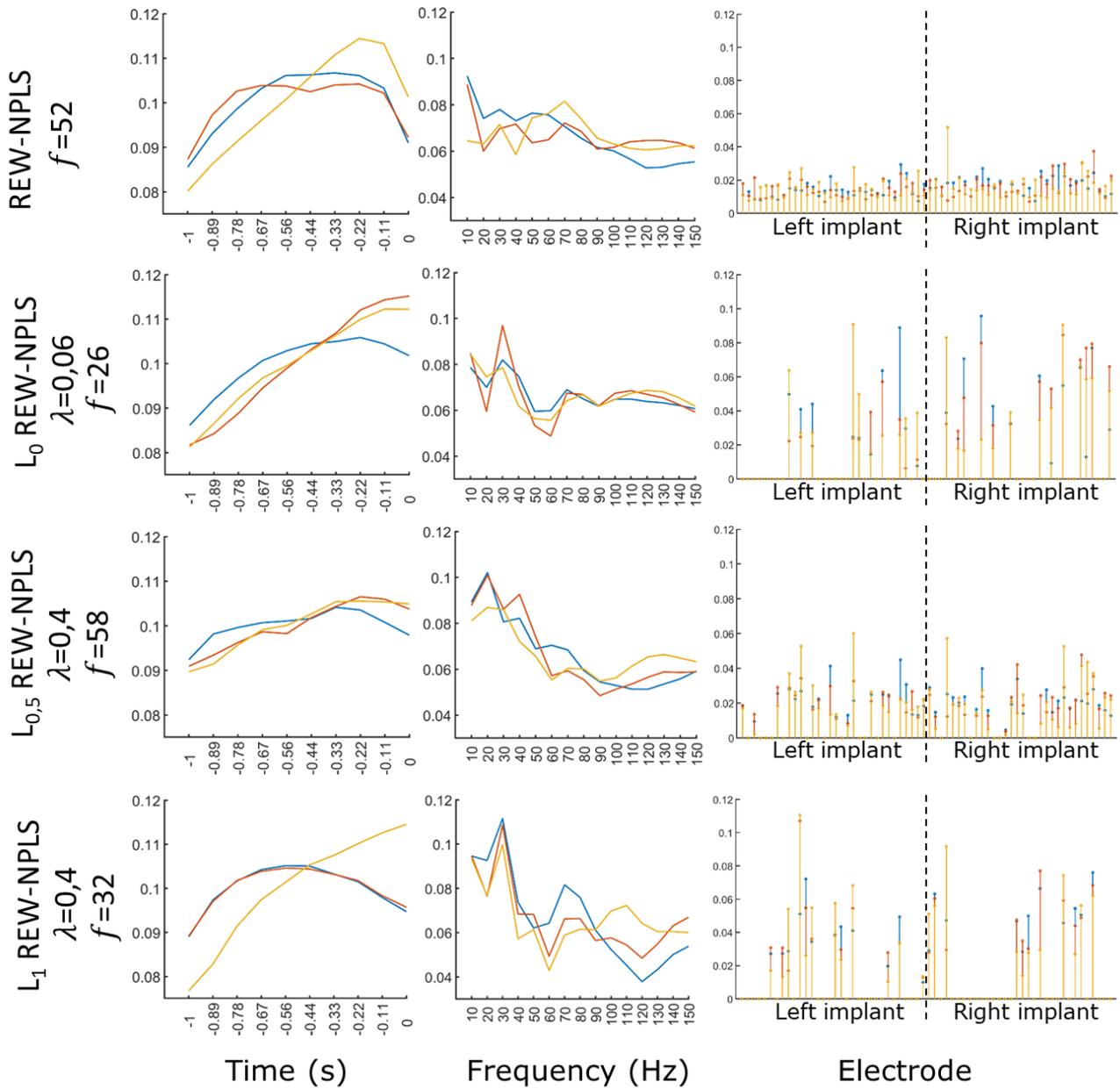

*Figure S2: Parameter weights of the Lp REW-NPLS and REW-NPLS models estimated offline in the left arm decoding study. Model parameter weights of the tested algorithms for 3D left arm decoding from the D dataset according to the spatial, frequency or temporal modalities. The parameter weights related to the $y_1, y_2$ and $y_3$ axis are represented using blue, orange and yellow lines respectively.*

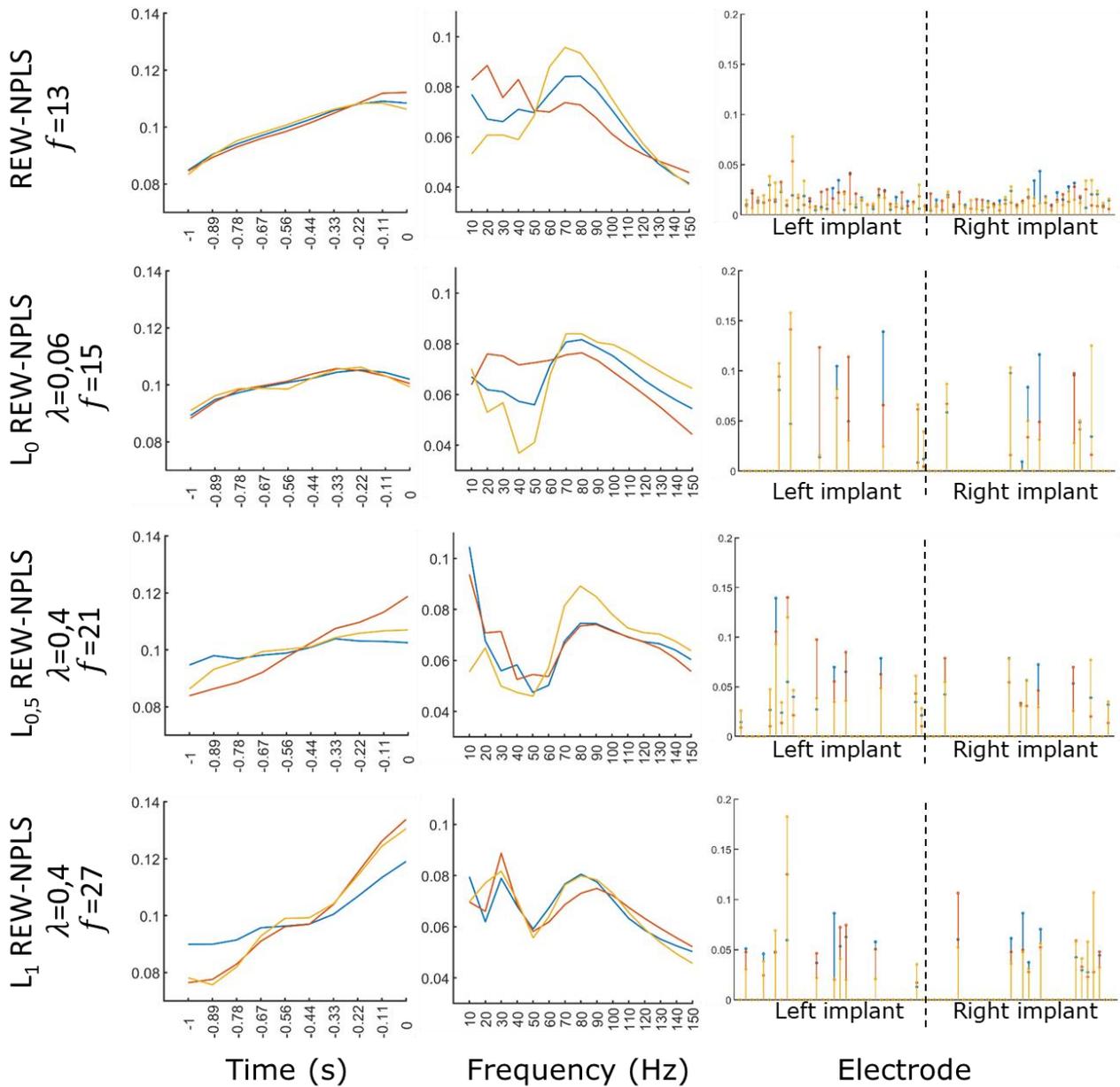

*Figure S3: Parameter weights of the Lp REW-NPLS and REW-NPLS models estimated offline in the right arm decoding study. Model parameter weights of the tested algorithms for 3D right arm decoding from the D dataset according to the spatial, frequency or temporal modalities. The parameter weights related to the $y_4$, $y_5$ and $y_6$ axis are represented using blue, orange and yellow lines respectively.*

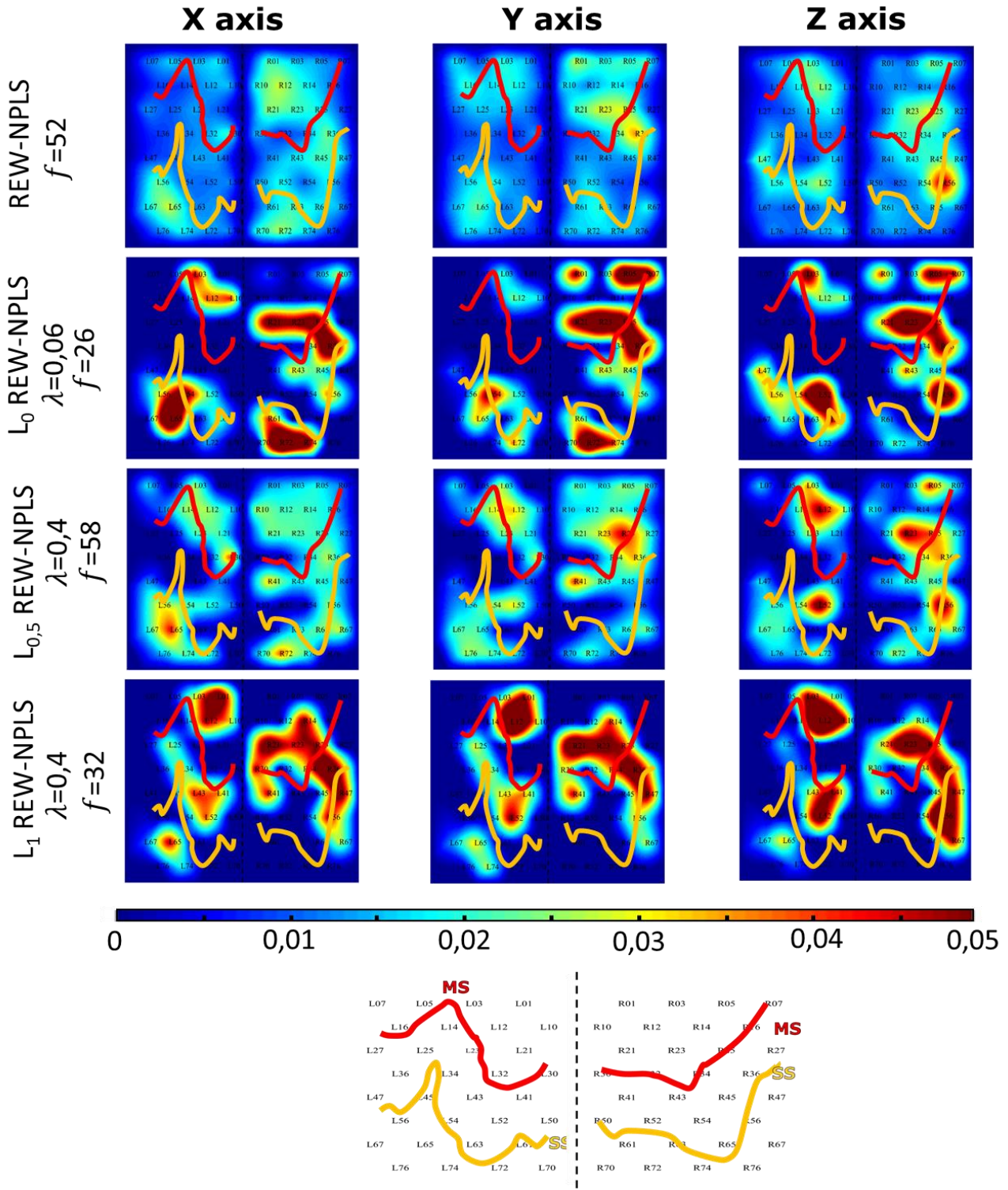

*Figure S4: 3D left hand decoding parameter weights of the three PREW-NPLS models projected on the spatial modality depending on the electrode location on the implant. The optimal latent space dimension f estimated during the Recursive-Validation procedure is used. The sensory sulcus (SS) and motor sulcus (MS) are represented in the spatial domain in yellow and red curves respectively.*

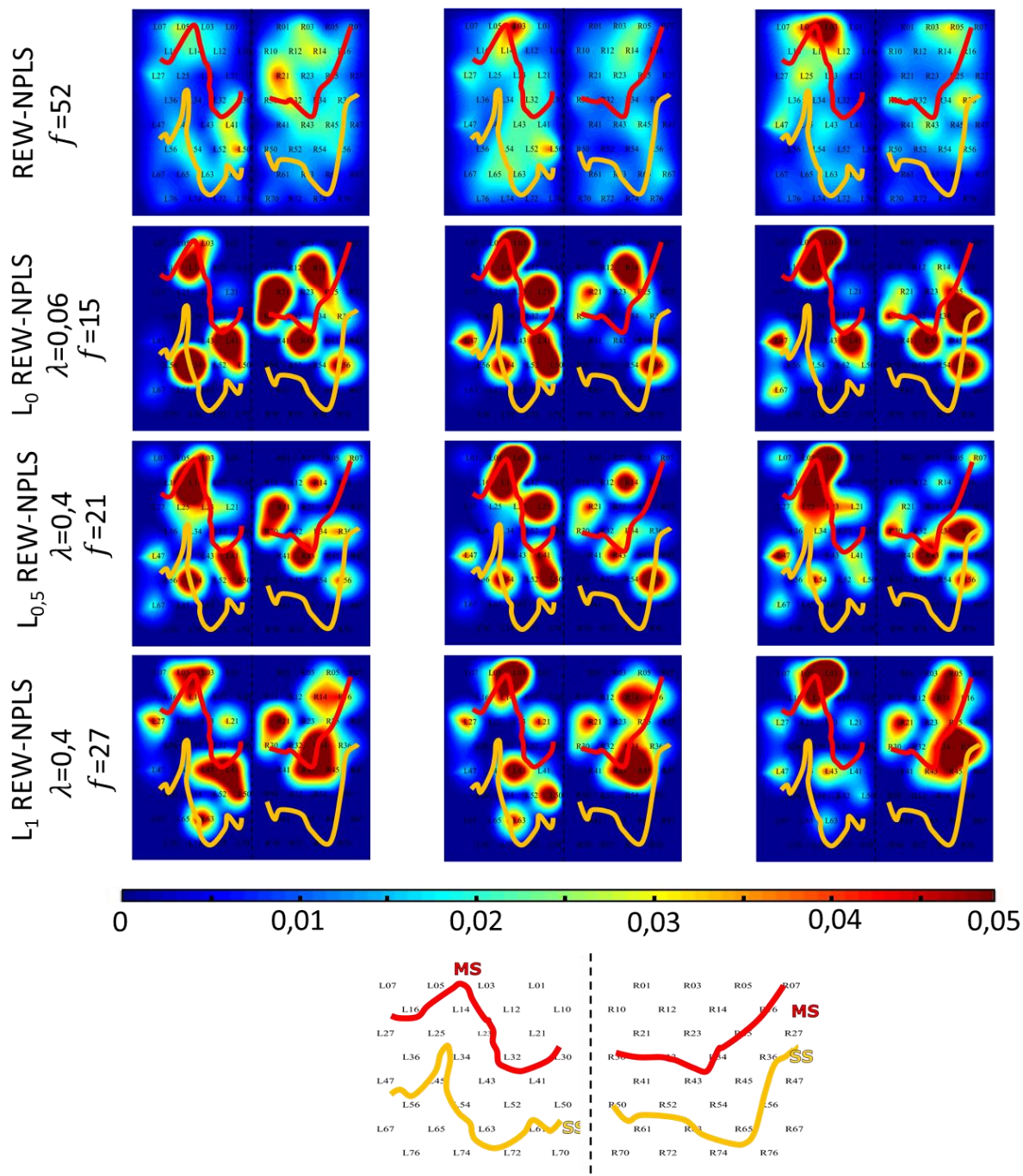

*Figure S5: 3D right hand decoding parameter weights of the three PREW-NPLS models projected on the spatial modality depending on the electrode location on the implant. The optimal latent space dimension f estimated during the Recursive-Validation procedure is used. The sensory sulcus (SS) and motor sulcus (MS) are represented in the spatial domain in yellow and red curves respectively.*

## Demonstration

Considering one iteration of the PARAFAC tensor decomposition algorithm of the tensor $\underline{\mathbf{V}}_u R^{I_1 \times I_2 \times I_3}$, the procedure estimated the projectors $\mathbf{w}^1, \mathbf{w}^2, \mathbf{w}^3$ such as:

$$\min_{\widehat{\underline{\mathbf{V}}}_u} \|\underline{\mathbf{V}}_u - \widehat{\underline{\mathbf{V}}}_u\|$$
$$\widehat{\underline{\mathbf{V}}}_u = \rho \, \mathbf{w}^1 \circ \mathbf{w}^2 \circ \mathbf{w}^3.$$
$$\|\mathbf{w}^1\| = \|\mathbf{w}^2\| = \|\mathbf{w}^3\| = 1.$$

Equally

$$\min_{\widehat{\underline{\mathbf{XY}}}_u} \|\underline{\mathbf{V}}_u - \widehat{\underline{\mathbf{V}}}_u\|^2 \tag{S1.1}$$
$$\widehat{\underline{\mathbf{V}}}_u = \rho \, \mathbf{w}^1 \circ \mathbf{w}^2 \circ \mathbf{w}^3.$$
$$\|\mathbf{w}^1\| = \|\mathbf{w}^2\| = \|\mathbf{w}^3\| = 1.$$

ALS is used to solve the task. It optimizes sequentially

$$\min_{\mathbf{w}^1} \left\| \underline{\mathbf{V}}_{u(1)} - \mathbf{w}^1 (\mathbf{w}^3 \otimes \mathbf{w}^2)^{\mathrm{T}} \right\|^2, \tag{S1.2}$$

$$\min_{\mathbf{w}^2} \left\| \mathbf{V}_{u(2)} - \mathbf{w}^2 (\mathbf{w}^3 \otimes \mathbf{w}^1)^{\mathrm{T}} \right\|^2, \tag{S1.3}$$

$$\min_{\mathbf{w}^3} \left\| \underline{\mathbf{V}}_{u(3)} - \mathbf{w}^3 (\mathbf{w}^2 \otimes \mathbf{w}^1)^{\mathrm{T}} \right\|^2 \tag{S1.4}$$

until convergence [1]. Let defined $\mathbf{w}^i_\rho = \mathbf{w}^i \rho$ with $i = 1,2,3$. The least Square (LS) solutions for each step are

$$\mathbf{w}^1_\rho = \frac{\underline{\mathbf{V}}_{u(1)} (\mathbf{w}^3 \otimes \mathbf{w}^2)}{\|\mathbf{w}^3 \otimes \mathbf{w}^2\|^2}, \tag{S1.5}$$

$$\mathbf{w}^2_\rho = \frac{\underline{\mathbf{V}}_{u(2)} (\mathbf{w}^3 \otimes \mathbf{w}^1)}{\|\mathbf{w}^3 \otimes \mathbf{w}^1\|^2}, \tag{S1.6}$$

$$\mathbf{w}^3_\rho = \frac{\underline{\mathbf{V}}_{u(3)} (\mathbf{w}^2 \otimes \mathbf{w}^1)}{\|\mathbf{w}^2 \otimes \mathbf{w}^1\|^2}. \tag{S1.7}$$

As $\|\mathbf{w}^i\|$, $i = 1,2,3$ are arbitrary values in (1.2)-(1.4), the resulted vectors $\mathbf{w}^1, \mathbf{w}^2, \mathbf{w}^3$ are normalized $\|\mathbf{w}^1\| = \|\mathbf{w}^2\| = \|\mathbf{w}^3\| = 1$ after convergence. Normalization allow the estimation of the parameter $\rho$. |

Let us denote the unfolded tensor $\underline{\mathbf{V}}_{(i)}$ with $\underline{\mathbf{V}}_{(i)} = \left(\mathbf{v}_1^1 | \ldots | \mathbf{v}_1^{I_1}\right) \in R^{I_1 \times I_2 I_3}$ where $\mathbf{v}_i^j$ are the columns of matrix $\mathbf{V}_i$. Taking into account that $(\mathbf{w}^2 \otimes \mathbf{w}^1)^{\mathrm{T}} \in R^{I_1 I_2}$, $(\mathbf{w}^3 \otimes \mathbf{w}^1)^{\mathrm{T}} \in R^{I_1 I_3}$ and $(\mathbf{w}^3 \otimes \mathbf{w}^2)^{\mathrm{T}} \in R^{I_2 I_3}$ are vectors, optimization tasks (S1.2)-(S1.4) are separated into element-wise optimization:

$$\min_{w_j^1} \|\mathbf{v}_1^j - w_j^1 (\mathbf{w}^3 \otimes \mathbf{w}^2)^{\mathrm{T}}\|^2, j = 1, \ldots I_1, \tag{S1.8}$$

$$\min_{w_j^2} \|\mathbf{v}_2^j - w_j^2 (\mathbf{w}^3 \otimes \mathbf{w}^1)^{\mathrm{T}}\|^2, j = 1, \ldots I_2, \tag{S1.9}$$

$$\min_{w_j^3} \|\mathbf{v}_3^j - w_j^3 (\mathbf{w}^2 \otimes \mathbf{w}^1)^{\mathrm{T}}\|^2, j = 1, \ldots I_3, \tag{S1.10}$$

where $w_j^i$ are elements of vectors $\mathbf{w}^1 = \left(w_1^1, \ldots, w_{I_1}^1\right)^T \in \mathbb{R}^{*I_1}$, $\mathbf{w}^2 = \left(w_1^2, \ldots, w_{I_2}^2\right)^T \in \mathbb{R}^{*I_2}$, and $\mathbf{w}^3 = \left(w_1^3, \ldots, w_{I_3}^3\right)^T \in \mathbb{R}^{*I_3}$. (1.5)-(1.7) may be written as

$$\left(w_j^1\right)_{LS} = \frac{\mathbf{v}_1^j (\mathbf{w}^3 \otimes \mathbf{w}^2)}{\|\mathbf{w}^3 \otimes \mathbf{w}^2\|^2}, j = 1, \ldots I_1 \tag{S1.11}$$

$$\left(w_j^2\right)_{LS} = \frac{\mathbf{v}_2^j (\mathbf{w}^3 \otimes \mathbf{w}^1)}{\|\mathbf{w}^3 \otimes \mathbf{w}^1\|^2}, j = 1, \ldots I_2 \tag{S1.12}$$

$$\left(w_j^3\right)_{LS} = \frac{v_3^j(\mathbf{w}^2 \otimes \mathbf{w}^1)}{\|\mathbf{w}^2 \otimes \mathbf{w}^1\|^2}, j = 1, \ldots I_3. \tag{S1.13}$$

In the invention, sparse promoting penalization using Lp (p=0,1/2, 1) norm/pseudo norms is proposed to integrate to cost function of REW-NPLS procedure to provide a group-wise sparse solutions, namely, solutions sparse by slices. Optimization task (S1.1) is replaced by optimization of the cost function penalized with Lp (p=0,1/2, 1) norm/pseudo norms

$$\min\|\underline{\mathbf{V}} - \widehat{\underline{\mathbf{V}}}\|^2 + P(\mathbf{w}^1, \mathbf{w}^2, \mathbf{w}^3), \tag{S1.14}$$

$$P(\mathbf{w}^1, \mathbf{w}^2, \mathbf{w}^3) = \lambda_1 \|\mathbf{w}^1\|_{q,\mathcal{L}_1} + \lambda_2 \|\mathbf{w}^2\|_{q,\mathcal{L}_2} + \lambda_3 \|\mathbf{w}^3\|_{q,\mathcal{L}_3},$$

$$\|\mathbf{w}^1\| = \|\mathbf{w}^2\| = \|\mathbf{w}^3\| = 1.$$

$\|\mathbf{w}^i\|_{p,\mathcal{L}_i}, p = 0, 1, 1/2, i = 1, 2, 3,$ is denoted as

$$\|\mathbf{w}^i\|_{0,\mathcal{L}_i} = \sum_{k \in \mathcal{L}_i} \left(1 - \delta_{0,w_k^i}\right),$$

$$\|\mathbf{w}^i\|_{1,\mathcal{L}_i} = \sum_{k \in \mathcal{L}_i} |w_k^i|,$$

$$\|\mathbf{w}^i\|_{\frac{1}{2},\mathcal{L}_i} = \sum_{k \in \mathcal{L}_i} \sqrt{|w_k^i|}.$$

Here, the regularization functions may only regularize a part of the indices defined by a set $\mathcal{L}_i \subset \{1,2,\ldots,I^i\}$, and protecting other elements of vector. $\mathcal{L}_i$ may vary depending on Rew NPLS iteration. $1 > \lambda_i > 0$ are regularization coefficients. A Kronecker delta $\delta_{0,w_k^i} = 1$ if $w_k^i = 0$, $\delta_{0,w_k^i} = 0$ otherwise.

The same ALS strategy (S1.2)-(S1.4) as used in conventional Rew NPLS is proposed to be apply for optimization (S1.14). ALS fixed all projectors except one at each step of the algorithm, leading to the three successive optimization tasks:

$$\min_{\mathbf{w}^1} \left(\|\underline{\mathbf{V}}_{u_{(1)}} - \mathbf{w}^1(\mathbf{w}^3 \otimes \mathbf{w}^2)^T\|^2 + \lambda_1 \|\mathbf{w}^1\|_{q,\mathcal{L}_1}\right),$$

$$\min_{\mathbf{w}^2} \left(\|\underline{\mathbf{V}}_{u_{(2)}} - \mathbf{w}^2(\mathbf{w}^3 \otimes \mathbf{w}^1)^T\|^2 + \lambda_2 \|\mathbf{w}^2\|_{q,\mathcal{L}_2}\right),$$

$$\min_{\mathbf{w}^3} \left(\|\underline{\mathbf{V}}_{u_{(3)}} - \mathbf{w}^3(\mathbf{w}^2 \otimes \mathbf{w}^1)^T\|^2 + \lambda_3 \|\mathbf{w}^3\|_{q,\mathcal{L}_3}\right).$$

The solution of non-regularized problem is used as initial approximation. Previously, similar penalized ALS was considered in [2] with whole regularized set of variables (no protection) and for L1 norm only. Moreover, the problem was solved using time consuming numerical optimization and was used offline only. In the current manuscript, more general case of Lp (p=0,1/2, 1) norm/pseudo norm penalization with possibility of variables protection which allow efficient integration to REW-NPLS algorithm is considered. Moreover, more efficient optimization procedure compatible with online real time applications is invented.

Firstly, we noted that contrary to non- regularized ALS (S1.2)-(S1.4), norms of projectors are not arbitrary parameters any more due to penalization terms. Therefore, the normalization of current estimate is added into to ALS optimization cycle.

$$\min_{\widetilde{\mathbf{w}}^1} \left(\|\underline{\mathbf{V}}_{u_{(1)}} - \widetilde{\mathbf{w}}^1(\mathbf{w}^3 \otimes \mathbf{w}^2)^T\|^2 + \lambda_1 \|\widetilde{\mathbf{w}}^1\|_{q,\mathcal{L}_1}\right), \quad \mathbf{w}^1 = \widetilde{\mathbf{w}}^1/\|\widetilde{\mathbf{w}}^1\| \tag{S1.15}$$

$$\min_{\widetilde{\mathbf{w}}^2} \left(\|\underline{\mathbf{V}}_{u_{(2)}} - \widetilde{\mathbf{w}}^2(\mathbf{w}^3 \otimes \mathbf{w}^1)^T\|^2 + \lambda_2 \|\widetilde{\mathbf{w}}^2\|_{q,\mathcal{L}_2}\right), \quad \mathbf{w}^2 = \widetilde{\mathbf{w}}^2/\|\widetilde{\mathbf{w}}^2\| \tag{S1.16}$$

$$\min_{\widetilde{\mathbf{w}}^3} \left(\|\underline{\mathbf{V}}_{u_{(3)}} - \widetilde{\mathbf{w}}^3(\mathbf{w}^2 \otimes \mathbf{w}^1)^T\|^2 + \lambda_3 \|\widetilde{\mathbf{w}}^3\|_{q,\mathcal{L}_3}\right) \quad \mathbf{w}^3 = \widetilde{\mathbf{w}}^3/\|\widetilde{\mathbf{w}}^3\|. \tag{S1.17}$$

Next, for faster computing, it can be noted that all considered regularization functions are decomposed as a sum of element-wise functions. Consequently, similarly to (S1.8)-(S1.10) optimization tasks (S1.15)-(S1.17) are split into element-wise optimization:

$$\min_{w_j^1} \left( \left\| \mathbf{v}_1^j - w_j^1 (\mathbf{w}^3 \otimes \mathbf{w}^2)^T \right\|^2 + \lambda_1 g_q(w_j^1) \right), j = 1, \ldots I_1 \tag{S1.18}$$

$$\min_{w_j^2} \left( \left\| \mathbf{v}_2^j - w_j^2 (\mathbf{w}^3 \otimes \mathbf{w}^1)^T \right\|^2 + \lambda_2 g_q(w_j^2) \right), j = 1, \ldots I_2 \tag{S1.19}$$

$$\min_{w_j^3} \left( \left\| \mathbf{v}_3^j - w_j^3 (\mathbf{w}^2 \otimes \mathbf{w}^1)^T \right\|^2 + \lambda_3 g_q(w_j^3) \right), j = 1, \ldots I_3 \tag{S1.20}$$

$$g_p(w_j^i) = \begin{cases} 1 - \delta_{0,w_j^i}, & \text{if } p = 0 \text{ and } w_j^i \in \mathcal{L}_i \\ |w_j^i|, & \text{if } p = 1 \text{ and } w_j^i \in \mathcal{L}_i \\ \sqrt{|w_j^i|} & \text{if } p = 1/2 \text{ and } w_j^i \in \mathcal{L}_i \\ 0 & \text{otherwise} \end{cases}$$

Let us consider particular the cases of $L_0$, $L_1$, $L_{1/2}$ norm penalizations.

### Penalization $L_0$

Let us consider one of optimization step, e.g. (S1.18) of ALS optimization in case of $L_0$ penalization:

$$FopL_0(w_j^1) = \|\mathbf{v}_1^j - w_j^1(\mathbf{w}^3 \otimes \mathbf{w}^2)^T\|^2 + \lambda_1\left(1 - \delta_{0,w_j^1}\right) \to min. \quad (1.21)$$

Here $\delta_{0,w_j^1} = 1$ if $w_j^1 = 0$, and $\delta_{0,w_j^1} = 0$ otherwise. Solution of (S1.21) is either $w_j^1 = 0$ or, if $w_j^1 \neq 0$, then $\delta_{0,w_j^1} = 0$,

$$FopL_0(w_j^1) = \|\mathbf{v}_1^j - w_j^1(\mathbf{w}^3 \otimes \mathbf{w}^2)^T\|^2 + \lambda_1 \to min,$$

$$\underset{w_j^1}{\mathrm{argmin}}\left(\|\mathbf{V}_1^j - w_j^1(\mathbf{w}^3 \otimes \mathbf{w}^2)^T\|^2 + \lambda_1\right) = \underset{w_j^1}{\mathrm{argmin}}\left(\|\mathbf{v}_1^j - w_j^1(\mathbf{w}^3 \otimes \mathbf{w}^2)^T\|^2\right)$$

The LS solution of non-penalized task is given by (S1.11): $(w_j^1)_{LS} = \frac{\mathbf{v}_1^j(\mathbf{w}^3 \otimes \mathbf{w}^2)}{\|\mathbf{w}^3 \otimes \mathbf{w}^2\|^2}$. In order to choose the solution of (S1.21) one may compare $FopL_0(O)$ and $FopL_0\left((w_j^1)_{LS}\right) = FopL_0\left(\frac{\mathbf{v}_1^j(\mathbf{w}^3 \otimes \mathbf{w}^2)}{\|\mathbf{w}^3 \otimes \mathbf{w}^2\|^2}\right)$. It is easy to see that $FopL_0(O) = (\mathbf{v}_1^j)^2$. The second candidate $FopL_0\left((w_j^1)_{LS}\right)$ takes the form

$$FopL_0\left((w_j^1)_{LS}\right) = FopL_0(O) - \|\mathbf{w}^3 \otimes \mathbf{w}^2\|^2\left((w_j^1)_{LS}\right)^2 + \lambda_1. \quad (S1.22)$$

From (S1.22), it follows that $L_0$ penalization correspond to hard thresholding of the least square solution $(w_j^1)_{LS}$:

$$(w_j^1)_{LS\_L_0} = \mathrm{argmin}\, FopL_0(w_j^1) = \begin{cases} 0, \text{if } i \in \mathcal{L}_1 \text{ and } (w_j^1)_{LS} \leq ThresholdL_0 \\ (w_j^1)_{LS} \text{ otherwise} \end{cases},$$

with $ThresholdL_0 = \frac{\sqrt{\lambda_1}}{\|\mathbf{w}^3 \otimes \mathbf{w}^2\|}$.

Other optimization steps allow similar consideration.

### Penalization $L_1$

An optimization step, e.g. (S1.18), of ALS optimization in case of $L_1$ penalization takes form

$$FopL_1(w_j^1) = \|\mathbf{v}_1^j - w_j^1(\mathbf{w}^3 \otimes \mathbf{w}^2)^T\|^2 + \lambda_1|w_j^1| \to min. \quad (S1.21)$$

In an inner-product space, the generalized Pythagorean theorem states that squared norm of sum of two orthogonal vectors is equal to sum of their squared norms [3]:

$$\|\mathbf{v}_1^j - w_j^1(\mathbf{w}^3 \otimes \mathbf{w}^2)^T\|^2 = \|\mathbf{v}_1^j - (w_j^1)_{LS}(\mathbf{w}^3 \otimes \mathbf{w}^2)^T\|^2 + \|(w_j^1)_{LS}(\mathbf{w}^3 \otimes \mathbf{w}^2)^T - w_j^1(\mathbf{w}^3 \otimes \mathbf{w}^2)^T\|^2,$$

due to orthogonality (see figure S6 for illustration). As only second term depend on the variable to optimize

$$\|(w_j^1)_{LS}(\mathbf{w}^3 \otimes \mathbf{w}^2)^T - w_j^1(\mathbf{w}^3 \otimes \mathbf{w}^2)^T\|^2 + \lambda_1|w_j^1| \to min. \quad (S1.22)$$

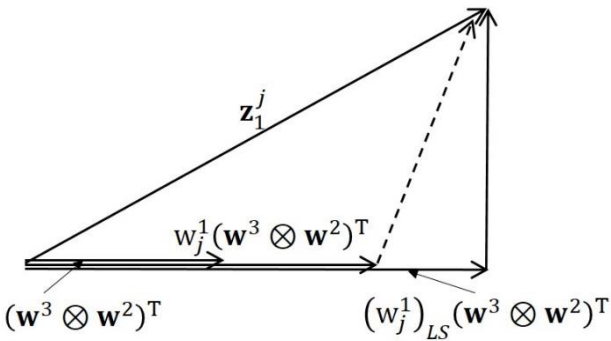

Figure S6.

Let us consider the case when scalar product of $\mathbf{v}_1^j$ and $(\mathbf{w}^3 \otimes \mathbf{w}^2)^T$ is non-negative, and $w_j^1 \geq 0$ (see figure) and the opposite case when $w_j^1 \leq 0$. In the first case

$$\left\|(w_j^1)_{LS}(\mathbf{w}^3 \otimes \mathbf{w}^2)^T - w_j^1(\mathbf{w}^3 \otimes \mathbf{w}^2)^T\right\|^2 + \lambda_1 w_j^1 \to min. \tag{S1.23}$$

As vectors $(w_j^1)_{LS}(\mathbf{w}^3 \otimes \mathbf{w}^2)^T$ and $w_j^1(\mathbf{w}^3 \otimes \mathbf{w}^2)^T$ are collinear and have the same direction, triangle inequality for norm degrade to equality

$$\left((w_j^1)_{LS}\|\mathbf{w}^3 \otimes \mathbf{w}^2\| - w_j^1\|\mathbf{w}^3 \otimes \mathbf{w}^2\|\right)^2 + \lambda_1 w_j^1 \to min.$$

After differentiation and taking into account that $w_j^1 \geq 0$,

$$w_j^1 = \begin{cases} 0, & \text{if } \lambda_1 \geq (w_j^1)_{LS}\|\mathbf{w}^3 \otimes \mathbf{w}^2\|^2 \\ \frac{(w_j^1)_{LS}\|\mathbf{w}^3 \otimes \mathbf{w}^2\|^2 - \lambda_1}{\|\mathbf{w}^3 \otimes \mathbf{w}^2\|^2} & \text{otherwise} \end{cases}.$$

Second case $w_j^1 < 0$ is considered similarly resulting in soft thresholding

$$(w_j^1)_{LS\_L_1} = argmin\, FopL_1(w_j^1) = \begin{cases} 0, & \text{if } j \in \mathcal{L}_1 \text{ and } (w_j^1)_{LS} \leq ThresholdL_1 \\ sign\left((w_j^1)_{LS}\right)\left(\left|(w_j^1)_{LS}\right| - ThresholdL_1\right) & \text{if } i \in \mathcal{L}_1 \text{ and } (w_j^1)_{LS} > ThresholdL_1 \\ (w_j^1)_{LS} & \text{otherwise} \end{cases},$$

$$ThresholdL_1 = \frac{\lambda_1}{\|\mathbf{w}^3 \otimes \mathbf{w}^2\|^2}.$$

Penalization $L_{0.5}$

An optimization step, e.g. (1.18), of ALS optimization in case of $L_{0.5}$ penalization takes form

$$FopL_{0.5}(w_j^1) = \left\|\mathbf{v}_1^j - w_j^1(\mathbf{w}^3 \otimes \mathbf{w}^2)^T\right\|^2 + \lambda_1\sqrt{|w_j^1|} \to min, \tag{S1.24}$$

$$(w_j^1)_{LS\_L_{0.5}} = argmin\, FopL_1(w_j^1).$$

Similarly to $L_1$ case:

$$\widetilde{FopL}_{0.5}(w_j^1) = \left((w_j^1)_{LS}\|\mathbf{w}^3 \otimes \mathbf{w}^2\| - w_j^1\|\mathbf{w}^3 \otimes \mathbf{w}^2\|\right)^2 + \lambda_1\sqrt{|w_j^1|}$$

$$= \|\mathbf{w}^3 \otimes \mathbf{w}^2\|^2\left((w_j^1)_{LS} - w_j^1\right)^2 + \lambda_1\sqrt{|w_j^1|} \to min.$$

The first case is scalar product of $\mathbf{v}_1^j$ and $(\mathbf{w}^3 \otimes \mathbf{w}^2)^T$ is non-negative, and $w_j^1 \geq 0$. Taking into account that $w_j^1 \leq (w_j^1)_{LS}$, the interval of interest is $[0, (w_j^1)_{LS}]$.

For $w_j^1 > 0$ the derivatives of cost function are

$$\left(\widetilde{FopL}_{0.5}(w_j^1)\right)' = 2\|\mathbf{w}^3 \otimes \mathbf{w}^2\|^2\left((w_j^1)_{LS} - w_j^1\right) + \frac{\lambda_1}{2\sqrt{w_j^1}}, \tag{S1.25}$$

$$\left(\widetilde{FopL}_{0.5}(w_j^1)\right)'' = 2\|\mathbf{w}^3 \otimes \mathbf{w}^2\|^2 - \frac{\lambda_1}{4(w_j^1)^{3/2}}.$$

Let us note that if $\left(\widetilde{FopL}_{0.5}(w_j^1)\right)' > 0$ for $w_j^1 \in [0, (w_j^1)_{LS}]$, $\widetilde{FopL}_{0.5}(w_j^1)$ is monotonically increasing function at the interval $[0, (w_j^1)_{LS}]$ and has minimum at $w_j^1 = 0$. First, let us study the cases when first derivative of the cost function is positive. Let us note that $\left(FopL_{0.5}(w_j^1)\right)''$ is monotonically increase. $\left(\widetilde{FopL}_{0.5}(w_j^1)\right)'' < 0$ for $w_j^1 \in \left[0, (w_j^1)_{LS}\right]$ if $\lambda_1 > 8\|\mathbf{w}^3 \otimes \mathbf{w}^2\|^2 \left((w_j^1)_{LS}\right)^{\frac{3}{2}}$. For negative second derivative first derivative is monotonically decrease. As at the end point $\left(\widetilde{FopL}_{0.5}\left((w_j^1)_{LS}\right)\right)' > 0$, derivative is positive for all interval of interest and optimal $(w_j^1)_{LS\_L_{0.5}} = 0$ (Figure S7A). Otherwise $\left(\widetilde{FopL}_{0.5}(w_j^1)\right)'' = 0$ at $\widehat{w}_j^1 = \frac{(\lambda_1)^{\frac{2}{3}}}{4\|\mathbf{w}^3 \otimes \mathbf{w}^2\|^{\frac{2}{3}}}$.

$\left(\widetilde{FopL}_{0.5}(w_j^1)\right)'$ decreases if $w_j^1 \in [0, \widehat{w}_j^1]$ and increases $w_j^1 \in \left[\widehat{w}_j^1, (w_j^1)_{LS}\right]$. Moreover $\left(\widetilde{FopL}_{0.5}(\widehat{w}_j^1)\right)' > 0$ if $\lambda_1 > \frac{8}{3\sqrt{3}}\|\mathbf{w}^3 \otimes \mathbf{w}^2\|^2 \left((w_j^1)_{LS}\right)^{\frac{3}{2}}$. In this case optimal solution is still $(w_j^1)_{LS\_L_{0.5}} = 0$ (Figure S7B).

If $\left(\widetilde{FopL}_{0.5}(\widehat{w}_j^1)\right)' < 0$ the cost function $\widetilde{FopL}_{0.5}(w_j^1)$ has 2 local extrema located in $[0, \widehat{w}_j^1]$ and $\left[\widehat{w}_j^1, (w_j^1)_{LS}\right]$ (Figure S7C). Only second one may correspond to the minimum of cost function (Figure S7C). It may be computed explicitly from $\left(\widetilde{FopL}_{0.5}(w_j^1)\right)' = 0$. Squared (1.25) may be viewed as cubic equation

$$x(1-x)^2 = C, \quad x = w_j^1 / (w_j^1)_{LS}, \quad C = \frac{\lambda_1^2}{16\|\mathbf{w}^3 \otimes \mathbf{w}^2\|^4 \left((w_j^1)_{LS}\right)^3} \tag{S1.26}$$

The solution of (1.26) in the interval of interest $w_j^1 \in \left[0, (w_j^1)_{LS}\right]$ ($x \in [0,1]$) exist in the case under considration (Figure S8) as $C \in [0, 4/27]$ if $\lambda_1 \leq \frac{8}{3\sqrt{3}}\|\mathbf{w}^3 \otimes \mathbf{w}^2\|^2 \left((w_j^1)_{LS}\right)^{\frac{3}{2}}$. By the properties of the function (Figure S7C), the minimum of cost function may be a biggest root of (1.26) in the interval [0; 1] denoted $x^*$. The minimum is attained at one of the points: 0 or $x^*(w_j^1)_{LS}$. Distinguishing between these options is straightforward. Summerising,

$$(w_j^1)_{LS\_L_{0.5}} = \arg\min FopL_{0.5}(w_j^1)$$
$$= \begin{cases} 0, & \text{if } j \in \mathcal{L}_1 \text{ and } (w_j^1)_{LS} \leq ThresholdL_{0.5} \\ \arg\min\left(FopL_{0.5}(0), FopL_{0.5}\left(x^* \cdot (w_j^1)_{LS}\right)\right), & \text{if } i \in \mathcal{L}_1 \text{ and } (w_j^1)_{LS} > ThresholdL_{0.5} \\ (w_j^1)_{LS} & \text{otherwise} \end{cases}$$

$ThresholdL_{0.5} = \frac{3}{4}\left(\frac{\lambda_1}{\|\mathbf{w}^3 \otimes \mathbf{w}^2\|^2}\right)^{\frac{2}{3}}$, $x^*$ is a biggest root of (1.26) in the interval [0; 1].

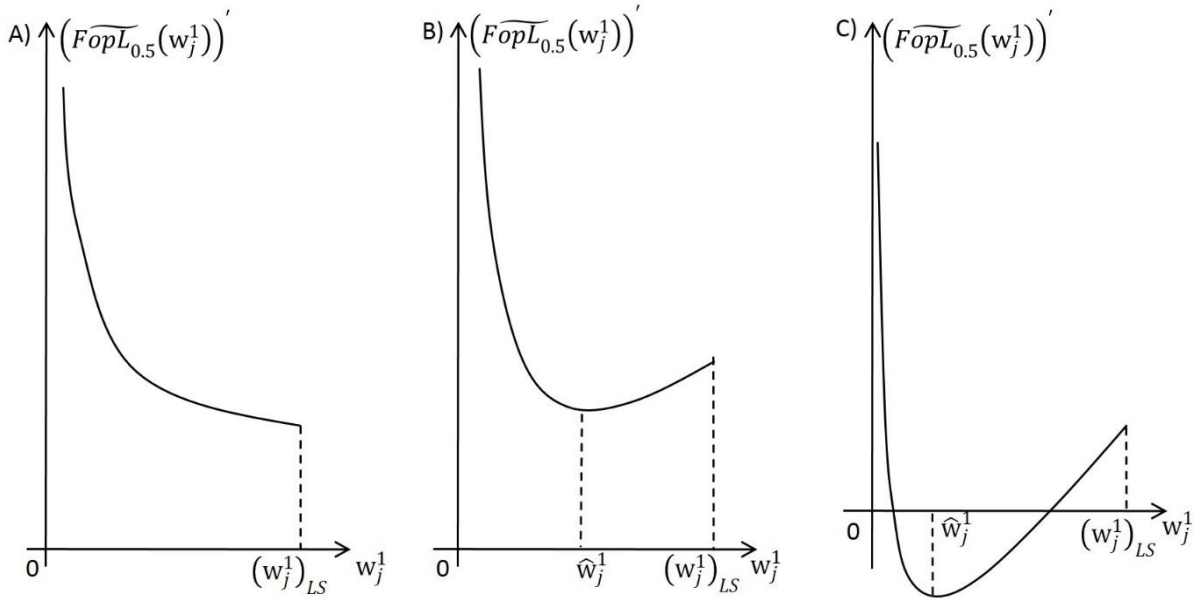

Figure S7.

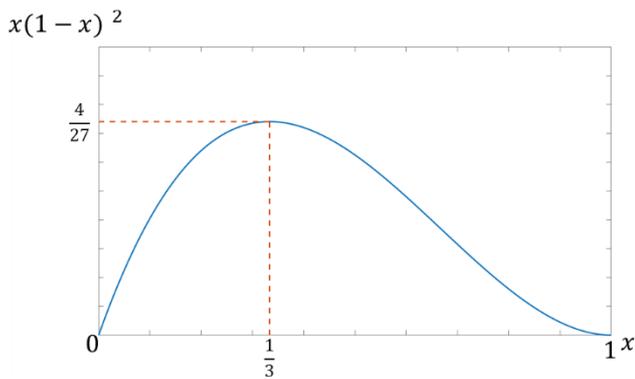

Figure S8

References of the demonstration section

[1] A. Uschmajew, « A new convergence proof for the higher-order power method and generalizations », *ArXiv14074586 Math*, janv. 2015, Consulté le: juin 25, 2020. [En ligne]. Disponible sur: http://arxiv.org/abs/1407.4586.

[2] Eliseyev A, Moro C, Faber J, Wyss A, Torres N, Mestais C, Benabid A L and Aksenova T 2012 L1-Penalized N-way PLS for subset of electrodes selection in BCI experiments *J. Neural Eng.* **9** 045010

[3] « Pythagorean theorem », *Wikipedia*. mai 26, 2020, Consulté le: juin 29, 2020. [En ligne]. Disponible sur: https://en.wikipedia.org/w/index.php?title=Pythagorean_theorem&oldid=958899723.